\documentclass[submission, Phys]{SciPost}

\hypersetup{
	colorlinks,
	linkcolor={red!50!black},
	citecolor={blue!50!black},
	urlcolor={blue!80!black}
}

\usepackage[bitstream-charter]{mathdesign}
\DeclareSymbolFont{usualmathcal}{OMS}{cmsy}{m}{n}
\DeclareSymbolFontAlphabet{\mathcal}{usualmathcal}

\usepackage[T1]{fontenc} 
\usepackage[textsize=footnotesize,textwidth=2.5cm]{todo notes}
\usepackage{cleveref}
\usepackage{caption}
\usepackage{subcaption}
\usepackage{physics}

\begin{document}
	
	\begin{center}{\Large \textbf{Entanglement Islands from Hilbert Space Reduction}}\end{center}
	
	\begin{center}
		Debarshi Basu\textsuperscript{1,2$\ast$},
		Qiang Wen\textsuperscript{1 $\dagger$} and Shangjie Zhou\textsuperscript{1,3$\ast$}
		
	\end{center}
	
	\begin{center}
		{\bf 1} Shing-Tung Yau Center and  School of Physics, Southeast University, Nanjing 210096, China
		\\
		{\bf 2} Indian Institute of Technology,
		Kanpur 208016, India\\
		{\bf 3}  School of Physics and Technology, Wuhan University, Wuhan, Hubei 430072, China
		
		$\ast$ These authors contribute equally to the paper
		\\
		$\dagger$ Corresponding to: { wenqiang@seu.edu.cn}
	\end{center}
	\begin{center}
		\today
	\end{center}
	
	
	\section*{Abstract}
	{\bf  { In this paper we propose a mechanism to generate entanglement islands in quantum systems from a purely quantum information perspective. More explicitly we show that, if we impose certain constraints on a quantum system by projecting out certain states in the Hilbert space, it is possible that for all the states remaining in the reduced Hilbert space, there exits subsets $I_a$ whose states are encoded in the states of another subset $\mathcal{R}_a$. Then the subsets $\{I_a\}$ are just the entanglement islands of the corresponding subsets $\{\mathcal{R}_a\}$. We call such a system \textit{self-encoded}, and find that the entanglement entropy in such systems should be calculated by a new \textit{island formula}. We give a comparison between our new island formula and \textit{island formula} in gravitational theories. Inspired by our mechanism, we propose a simulation of the AdS/BCFT correspondence and the island phases in this context via a holographic CFT$_2$ with a special Weyl transformation. }
	}

	\vspace{10pt}
	\noindent\rule{\textwidth}{1pt}
	\tableofcontents\thispagestyle{fancy}
	\noindent\rule{\textwidth}{1pt}
	\vspace{10pt}
	
	\section{Introduction}
The information paradox for evaporating black holes \cite{Hawking:1976ra,Page:1993wv,Page:2013dx,Mathur:2009hf,Almheiri:2012rt} is one of the most important mysteries in our understanding of nature. It was expected that finding a solution to the information paradox could lead us to a window to understand the quantum theory of gravity. 
The AdS/CFT correspondence \cite{Maldacena:1997re,Gubser:1998bc,Witten:1998qj} that equates asymptotically AdS gravitational theories to certain conformal field theories with large central charges and strong coupling, provides us a framework to study the quantum aspects of gravity. 
This is called the holographic nature of gravity and is believed to be a general property for gravitational theories, which strongly indicates that the quantum theory of gravity should be manifestly unitary. Nevertheless, a concrete understanding of how the information is preserved during the black hole evaporation is not obvious at all.
A major breakthrough on this problem is based on the study of quantum entanglement structure of the holographic field theories. 
In the context of AdS/CFT, the Ryu-Takayanagi (RT) formula \cite{Ryu:2006bv,Ryu:2006ef,Lewkowycz:2013nqa} relates the entanglement entropy of any subregion in the boundary CFT to certain co-dimension two minimal (extremal) surfaces in the bulk which are homologous to the corresponding boundary subregion. 
This formula was further refined to the quantum extremal surface (QES) formula which included the quantum correction from bulk fields \cite{Faulkner:2013ana,Engelhardt:2014gca}\footnote{See \cite{Wall:2012uf,Dong:2016hjy,Akers:2019lzs,Dong:2017xht} for similar refinements of the covariant holographic entanglement entropy proposal in \cite{Hubeny:2007xt}.}. 

The authors of Ref. \cite{Penington:2019npb,Almheiri:2019psf,Rozali:2019day,Chen:2019uhq} applied the QES prescription to compute the entanglement entropy of the radiation from an evaporating black hole after the Page time. 
Remarkably they found that, the result deviates from Hawking's calculations and is consistent with unitary evolution. 
These computations further inspired the proposal of the so-called ``island formula'' \cite{Almheiri:2019hni, Almheiri:2019psy, Almheiri:2019yqk,Almheiri:2019qdq,Penington:2019kki}, which is claimed to be the formula to compute the entanglement entropy for regions in gravitational theories.  
The island formula has been extensively studied in configurations consisting of a system on a fixed spacetime background (or non-gravitational system) and a system with dynamical gravity (or in the context of AdS/BCFT\cite{Takayanagi:2011zk}). 
The two systems are glued together at some surface with transparent boundary conditions for matter fields, and hence the radiation from a black hole in the gravitating region can enter the non-gravitational system freely. In this setup the non-gravitational system plays the role of a reservoir that absorbs the Hawking radiation.

{ 
	In the following, we will encounter density matrices defined and their associated von-Neumann entropy $S(\rho)$ for different purposes due to the gravitational effects. It will be useful to clarify their differences using different notations, which are listed below,
	\begin{itemize}
		\item $\rho_{\mathcal{R}}$: the density matrix for any region $\mathcal{R}$ in the ``full quantum theory'' including quantum gravity effects for which we do not have the exact description yet;
		
		\item $\rho^{b}_{\mathcal{R}}$: the density matrix for any region $\mathcal{R}$ by tracing out all the degrees of freedom outside ${\mathcal{R}}$ in semi-classical \textit{effective theory} description, which is a quantum field theory defined on a fixed (curved) spacetime. Although the spacetime could be curved, in this paper we classify this description as non-gravitational since the gravitational fluctuations are omitted. The associated von Neumann entropy $S(\rho^{b}_{\mathcal{R}})$ is also called the bulk entanglement entropy.
		
		\item $\tilde{\rho}_{\mathcal{R}}$: the density matrix for any region $\mathcal{R}$ computed by tracing out the degrees of freedom outside $\mathcal{R}$. This definition is usually well-defined and equals to $\rho_{\mathcal{R}}$ when $\mathcal{R}$ is settled in a non-gravitational system. Nevertheless, it is just a intermediate concept in the computation of the fine-grained or entanglement entropy \cite{Faulkner:2013ana,Engelhardt:2014gca} when $\mathcal{R}$ is settled in a gravitational background. Here we naively define
		\begin{align}\label{trho}
			S(\tilde{\rho}_{\mathcal{R}})=\frac{Area(X)}{4G}+S(\rho^{b}(\mathcal{R}))\,,
		\end{align}
		where $X$ is the boundary surface of $\mathcal{R}$ and the area term comes from gravitational fluctuations. Then the entanglement entropy for regions in gravitational backgrounds are given by extremizing the above formula with respect to $\mathcal{R}$ (or $X$). In other words, when generalizing $\tilde{\rho}_{\mathcal{R}}$ from non-gravitational systems to gravitational backgrounds, $\tilde{\rho}_{\mathcal{R}}$ is only considered to be a well-defined density matrix for $\mathcal{R}$ when $\mathcal{R}$ is bounded by a quantum extremal surfaces (QES), in the sense that the computation of the fine-grained entropy can be described by the gravitational replica trick with twist operators inserted at the boundary of $\mathcal{R}$. Of course, for an arbitrary $\mathcal{R}$ in \eqref{trho}, $\tilde{\rho}_{\mathcal{R}}$ is not a density matrix, and $S(\tilde{\rho}_{\mathcal{R}})$ is defined by \eqref{trho} with no further physical interpretation as a entanglement entropy.    	   	
	\end{itemize} 
	All the above density matrices reduce to the standard one in quantum information theory when applied to non-gravitational systems with a factorizable Hilbert space.
	
	Let us first review Hawking's computation for the entanglement entropy of the Hawking radiation. We denote the reservoir that collects the Hawking radiation as $\mathcal{R}$, and it is the region far away from the black hole thus can be considered as a non-gravitational theory on a fixed spacetime. Since there is no gravitational fluctuations in $\mathcal{R}$, the area term in \eqref{trho} does not appear and we have $S{(\tilde{\rho}_{\mathcal{R}})}=S{(\rho^{b}_{\mathcal{R}})}$. In Hawking's calculation, it was believed that the semi-classical description is valid almost everywhere except the near singularity region (which is a much larger area than the reservoir $\mathcal{R}$) before the complete evaporation of the black hole. Furthermore, assuming locality for the semi-classical description, it was taken for granted that, the degrees of freedom at different sites on a Cauchy surface where the \textit{effective theory} lives should be independent from each other, hence the Hilbert space enjoys the factorization property, hence Hawking's calculation follows a standard definition of the reduced density matrix in quantum information. In other words, it was expected that $\rho_{\mathcal{R}}=\tilde{\rho}_{\mathcal{R}}=\rho^{b}_{\mathcal{R}}$, hence \cite{Hawking:1976ra}, 
	\begin{align}\label{Hawkingc}
		Hawking's~calculation:\qquad S(\rho_{\mathcal{R}})=S(\tilde{\rho}_{\mathcal{R}})=S(\rho^{b}_\mathcal{R})\,.
	\end{align}
	This calculation gives a monotonically increasing entanglement entropy for the Hawking radiation, until the complete evaporation of the black hole.
	
	\begin{figure}[ht]
		\centering
		\includegraphics[scale=0.6]{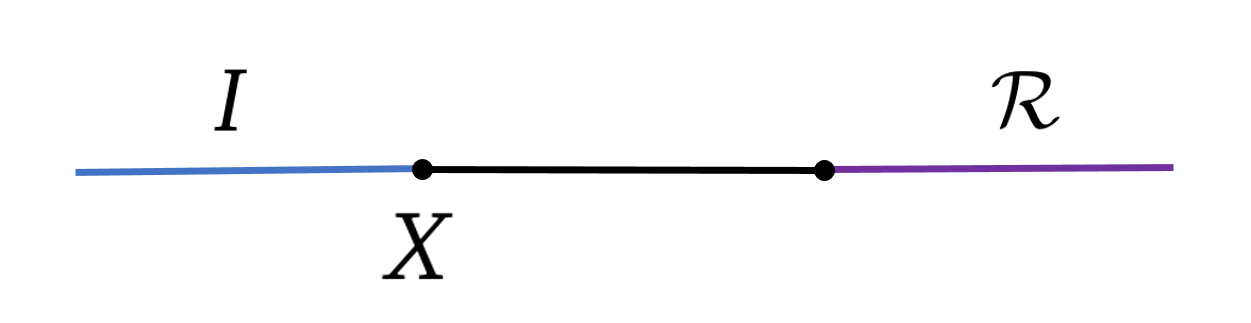}
		\caption{Schematics of the quantum extremal surface $X$ for a subregion $\mathcal{R}$ in a quantum field theory coupled to semiclassical gravity. The region $I$ in the gravitational background bounded by $X$ is the island corresponding to the region $\mathcal{R}$ in non-gravitational background.}
		\label{fig:island}
	\end{figure}
	
	Nevertheless, unitarity requires the entanglement entropy for the Hawking radiation to follow the Page curve \cite{Page:1993wv}, which starts to decrease after the Page time (when the black hole has roughly evaporated half of its mass), hence Hawking's calculation \eqref{Hawkingc} is not consistent with unitarity after the Page time. Also the Mathur/AMPS puzzle \cite{Mathur:2009hf,Almheiri:2012rt} arises after the Page time near the horizon, where the effective field theory description was believed to be valid. The island formula could be the answer to these problems, as it considers the gravitational fluctuations outside the reservoir $\mathcal{R}$. It claims that, when computing $S(\rho_\mathcal{R})$ we should not only consider the degrees of freedom inside $\mathcal{R}$, but also the degrees of freedom in any gravitational region $I$ outside $\mathcal{R}$. 
	More explicitly, $S(\rho_{\mathcal{R}})$ is calculated by the entanglement entropy formula \cite{Faulkner:2013ana,Engelhardt:2014gca} with the consideration of including certain island regions \cite{Almheiri:2019hni, Almheiri:2019psf, Almheiri:2019psy, Almheiri:2019yqk, Almheiri:2019qdq}:
	\begin{align}\label{islandformula1}
		Island~formula~\uppercase\expandafter{\romannumeral1}:\qquad S(\rho_{\mathcal{R}})=&\text{min}\left\{\text{ext}_{X} S(\tilde{\rho}_{I\cup \mathcal{R}})\right\}
		\cr
		=&\text{min}\left\{\text{ext}_{X}\left[\frac{Area(X)}{4G}+S(\rho^b_{I\cup \mathcal{R}})\right]\right\}\,,
	\end{align}
	Here $X$ is the boundary of $I$, and the above prescription means to extremize a generalized entropy-like functional over all possible $I$, then minimize over all extrema. The $I$ that solves the optimization problem in this formula is called the ``entanglement island'' of $\mathcal{R}$. In the first line, $\tilde{\rho}_{I\cup \mathcal{R}}$ is the intermediate reduced density matrix for $I\cup\mathcal{R}$ calculated by \eqref{trho}. Note that, the area term for the boundary surface of $\mathcal{R}$ does not appear as $\mathcal{R}$ is the non-gravitational reservoir. 
	
	Between the island formula \eqref{islandformula1} and the Hawking's formula \eqref{Hawkingc}, one should choose the one that gives the smaller $S(\rho_{\mathcal{R}})$. Remarkably, after the Page time \eqref{islandformula1} gives smaller $S(\rho_{\mathcal{R}})$, which follows the Page curve. We call the system turns to the island phase after the Page time. This implies that, in the island phase the true reduced density matrix $\rho_{\mathcal{R}}$ may not be given by the standard definition in quantum information, which is tracing out the degrees of freedom outside $\mathcal{R}$, i.e.
	\begin{align}\label{rdmqft}
	Island~phases:\qquad \rho_{\mathcal{R}}\neq \tilde{\rho}_{\mathcal{R}}\,.
	\end{align}

	The doubly holographic set-up discussed in \cite{Almheiri:2019hni} provides a special framework to explain the origin of the island formula. In this scenario, the quantum field theory describing the Hawking radiation is assumed to be holographic. The corresponding bulk dual gravitational theory (Fig.\ref{fig:double-holography1}) has a lower dimensional effective description (Fig.\ref{fig:double-holography2}) in terms of the radiation bath coupled to a AdS$_2$ gravitational theory where the entanglement islands could emerge. Furthermore, when the AdS$_2$ gravity part is holographically dual to a (0+1)-dimensional quantum dot, we arrived at the third picture of the same configuration, which is called the fundamental description (Fig.\ref{fig:double-holography3}). In the doubly holographic framework, the entanglement entropy of a subsystem in the radiation bath is computed through the usual (H)RT formula \cite{Ryu:2006bv,Hubeny:2007xt} which is equivalent to the island prescription in the lower dimensional effective description (Fig.\ref{fig:double-holography}). The doubly holographic setup naturally encapsulates the idea of the island in the black hole interior being encoded in the entanglement wedge of the radiation. Moreover, in a more general set-up without assuming holography, the island formula has been derived via gravitational path integrals where wormholes are allowed to exist as new saddles (called the replica wormholes) when calculating the partition function on the replica manifold \cite{Almheiri:2019qdq,Penington:2019kki}. For a subset of relevant works that may be related to this paper, see \cite{Chen:2019iro,Chen:2020wiq,Chen:2020hmv,Hernandez:2020nem,Grimaldi:2022suv, Akal:2020twv,Deng:2020ent, Anous:2022wqh, Geng:2020qvw, Karlsson:2020uga, Raju:2020smc, Raju:2021lwh, Laddha:2020kvp, Geng:2020fxl,Geng:2021hlu,Alishahiha:2020qza,HosseiniMansoori:2022hok,Karch:2022rvr,Uhlemann:2021nhu,Krishnan:2020oun,Krishnan:2020fer,Ghosh:2021axl}. Also, see \cite{Almheiri:2020cfm,Bousso:2022ntt} for a detailed review on this topic\footnote{On the other hand, there are still important criticisms \cite{Geng:2020qvw, Karlsson:2020uga, Raju:2020smc, Raju:2021lwh, Laddha:2020kvp, Geng:2020fxl,Geng:2021hlu} for the \textit{Island formula \uppercase\expandafter{\romannumeral1}} which remain to be properly addressed (see also \cite{Miao:2022mdx,Miao:2023unv} for alternative viewpoint).}. 
	\begin{figure}
		\centering
		\begin{subfigure}[t]{0.3\textwidth}
			\centering
			\includegraphics[width=\textwidth]{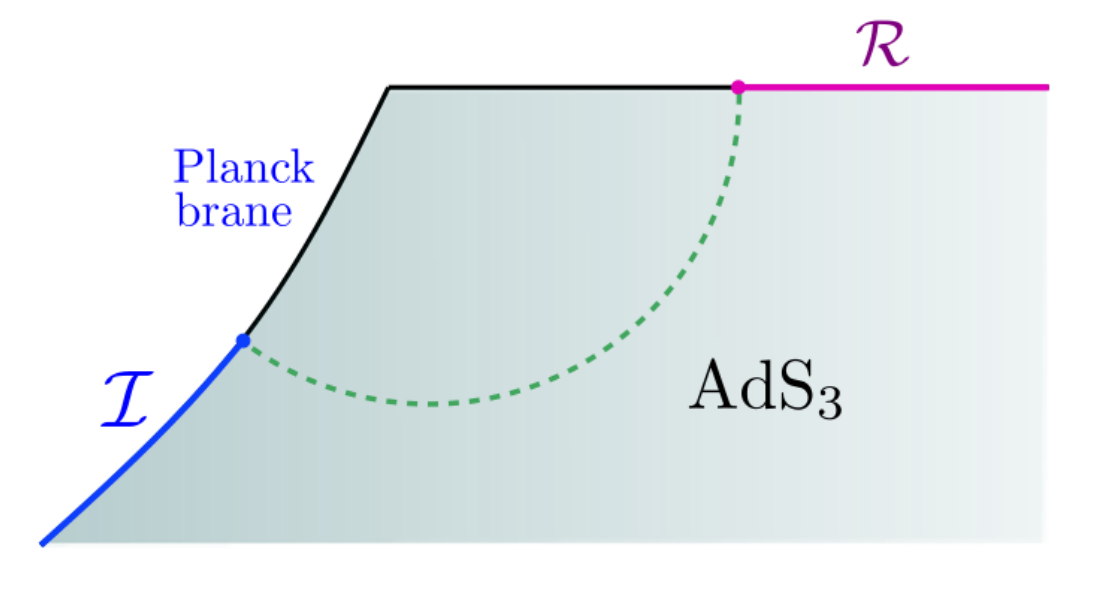}
			\caption{Double holography description: AdS$_3$ spacetime truncated by an end-of-the-world Planck brane}
			\label{fig:double-holography1}
		\end{subfigure}
		\hfill
		\begin{subfigure}[t]{0.3\textwidth}
			\centering
			\includegraphics[width=\textwidth]{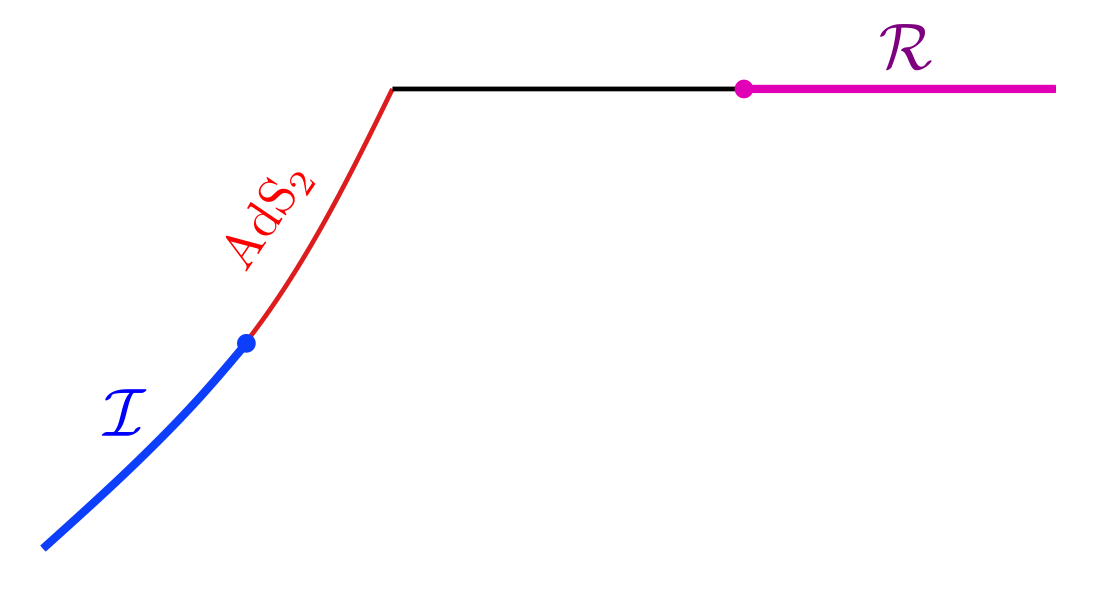}
			\caption{Effective theory description: CFT$_2$ on a half line coupled to semi-classical gravity on the AdS$_2$ brane.}
			\label{fig:double-holography2}
		\end{subfigure}
		\hfill
		\begin{subfigure}[t]{0.3\textwidth}
			\centering
			\includegraphics[width=\textwidth]{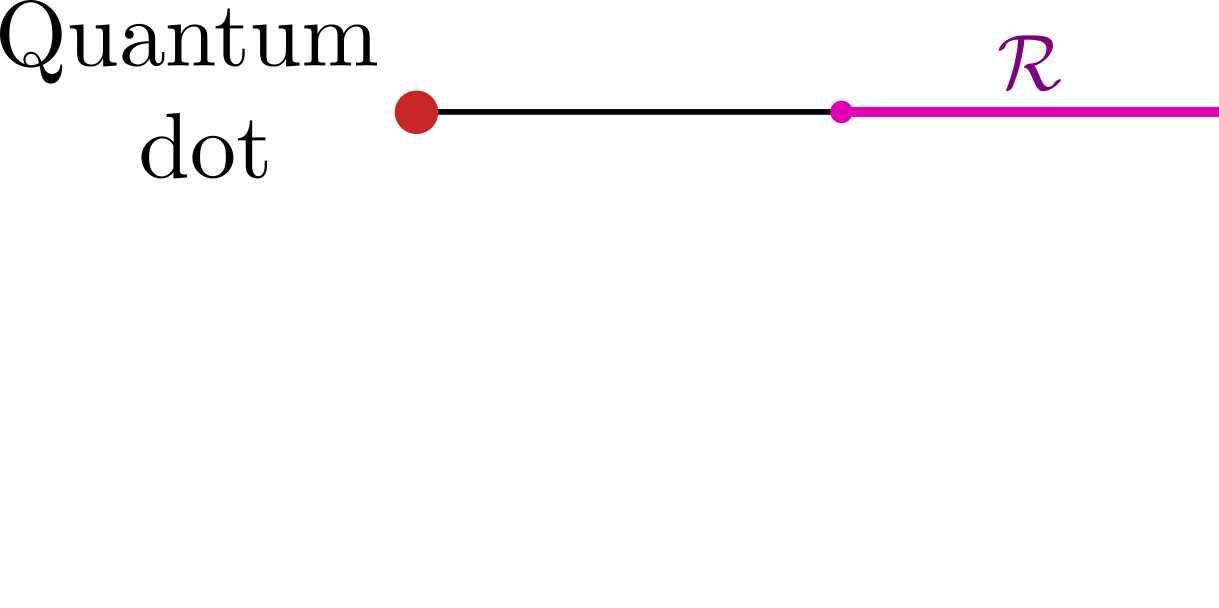}
			\caption{Fundamental description: CFT$_2$ on a half line coupled to a quantum dot.}
			\label{fig:double-holography3}
		\end{subfigure}
		\caption{Three different descriptions for the same configuration.}
		\label{fig:double-holography}
	\end{figure}
	
	In all the configurations we have reviewed, gravitation plays a crucial role for the emergence of islands. It is tempting to believe that entanglement islands only exist in gravitational backgrounds. Although the island formula for $S({\rho_{\mathcal{R}}})$ reproduces the Page curve, it indicates that when $\mathcal{R}$ admits an island $I$, the exact density matrix $\rho_{\mathcal{R}}$ is not calculated by \eqref{rdmqft} even $\mathcal{R}$ is settled in a non-gravitational background, rather it is seems to be calculated by
	\begin{align}\label{rdmet}
		\rho_{\mathcal{R}}=\tilde{\rho}_{\mathcal{R}\cup I}, \qquad \textit{I is the entanglement island of $\mathcal{R}$}\,.
	\end{align}
	Here the optimization is missing as $I$ is the entanglement island bounded by the QES $X$, which is already the solution of the optimization problem in \eqref{islandformula1}.
}

This is quite surprising and counter-intuitive to our standard understanding of fundamental quantum information. The above discussion lead us to the following question. Is gravitation essential for the emergence of entanglement islands? Or, is it possible to understand the Island formula from a purely quantum information perspective? These are fundamental questions and answering them will be crucial to get a deeper understanding of the entanglement islands. Furthermore, if the island formula can be understood from a purely quantum information perspective\footnote{See \cite{Renner:2021qbe,Wang:2021afl,Akers:2022qdl,Almheiri:2021jwq} for examples which, in some sense, also attempted to study entanglement islands in quantum information without gravitation.}, it may lead us to a new research field of quantum information, where we may find a way to create entanglement islands in the lab and discuss how to use them. 

In this paper we try to understand the above questions. In short we are going to study the reduced density matrices for quantum systems in a reduced Hilbert space, where the state of space-likely separated degrees of freedom becomes dependent on each other and the factorization property does not hold anymore. We will get a new formula to compute the entanglement entropy for regions in the reduced Hilbert space, which looks very similar to the island formula \eqref{islandformula1}. This will give us new inspiration on understanding the \textit{Island formula I} from a pure quantum information perspective.  More explicitly, in section \ref{sec2}, we propose a mechanism for the emergence of the entanglement islands in quantum systems, and derive the formula, which we call the \textit{Island formula II}, to compute entanglement entropy in systems under this mechanism. In section \ref{sec3} we compare the \textit{Island formula II} with the \textit{Island formula I} \eqref{islandformula1} in gravitational systems. In section \ref{sec:4-examples}, we propose a simulation for the AdS/BCFT configurations via holographic CFT$_2$ under a special Weyl transformation. In section \ref{sec:5-discussions}, we summarize our results and discuss their implications.

\section{Entanglement islands from Hilbert space reduction}\label{sec2}
\label{sec:3-Hilbert-reduction}
\subsection{Self-encoded quantum systems induced by constraints}
For a generic quantum system when the Hilbert space of the total system is properly reduced following certain constraints, { the entanglement islands $I_a$ for regions $\mathcal{R}_a$ could emerge in a natural way. In the following, we list the key features for the constraints that induce entanglement islands.
	\begin{itemize}
		\item \textit{Firstly, the constraints should be understood as projecting out certain states in the Hilbert space such that, for all the states remaining in the reduced Hilbert space, mappings from the states of a set of subregions $\left\{ \mathcal{R}_{a}\right\}$ to the state of another set of subregions $\left\{ I_{a} \right\}$ will emerge, 
			\begin{align}\label{codingrelations}
				\ket{i}_{I_{a}}=f_{\mathcal{R}_{a}}(\ket{j}_{\mathcal{R}_{a}})\,,
			\end{align}
			which we call the coding relations. {  More explicitly, let us begin with a quantum system with two sites $I$ and $\mathcal{R}$, hence a general basis of the Hilbert space can be written as $\ket{i}_{I}\ket{j}_{\mathcal{R}}$. For the basis with $j=0$, if we project out all the states $\ket{i}_{I}\ket{0}_{\mathcal{R}}$ with $ i\neq 0$, then in the reduced Hilbert space if you grab a state and read the $\mathcal{R}$ factor to be $\ket{0}_{\mathcal{R}}$, you will immediately notice that the $I$ factor of this state should be $\ket{0}_{I}$. The reason is that the states with other possible $i$ are not in the reduced Hilbert space. This effectively gives a mapping between the state of $\mathcal{R}$ and the state of $I$, which is the coding relation $\ket{0}_{I}=f_{\mathcal{R}}(\ket{0}_{\mathcal{R}})$. Of course we can perform more complicated Hilbert space reduction such that, given the state of $\mathcal{R}_{a}$ one can determine the state of $I_{a}$ through the coding relation. The coding relations can be classified in two levels, which include the 1) the region correspondence between $\{\mathcal{R}_{a},I_{a}\}$ and 2) the state correspondence between $\{\ket{j}_{\mathcal{R}_{a}},\ket{i}_{I_{a}}\}$ for a given pair of $\{\mathcal{R}_{a},I_{a}\}$.}}
		
		\item \textit{Secondly, the constraints are not only imposed on the states, but also on the Hilbert space of the system. In other words, under time evolution or the action of any allowed operators, the state of the system should always remain in the reduced Hilbert space, rather than the original full Hilbert space. Operators that drive the system to a state outside the reduced Hilbert space are not consistent with the constraints.}
	\end{itemize}
	We call systems satisfying such kind of constraints the \textit{self-encoded systems}. As was described above, the constraints are highly non-trivial and result in highly non-local effects to the system, which destroy the independence between spacelike-separated degrees of freedom on a Cauchy surface. 	
	This will essentially change the way we calculate the reduced density matrix and related information quantities like entanglement entropy. We will show that the new formula to calculate the entanglement entropy in self-encoded systems is closely related to the island formula \eqref{islandformula1}.}

Before proceeding, we would like to briefly comment on the nature of the quantum channels incorporating the coding relations. By virtue of being a quantum channel, such a coding relation must be a completely positive trace-preserving (CPTP) map. This ensures that the coding relation maps a density matrix (with unit trace) to another density matrix even in the presence of an environment. An example of such a coding relation is given by the so called Petz recovery channel \cite{1993QuantumEA}, which preserves relative entropies.

\subsection{The simplest case of two spins\label{subsec:two_spins}}
Now, we give a explicit description of our mechanism for a system with the simplest coding relation, and explicitly show how we should compute the reduced density matrix and the entanglement entropy in the reduced Hilbert space. For brevity we consider static systems in two-dimensional spacetime with time reflection symmetry. We divide the system into three non-overlapping subsets $\mathcal{R}\cup I\cup B$\footnote{The denotations are chosen to match the black hole configurations \cite{Almheiri:2020cfm}, where $\mathcal{R}$, $I$ and $B$ are analogues of the black hole radiation in the non-gravitational reservoir, the island in black hole interior and the black hole degrees of freedom respectively. }.

\begin{figure}[ht]
	\centering
	\includegraphics[scale=0.4]{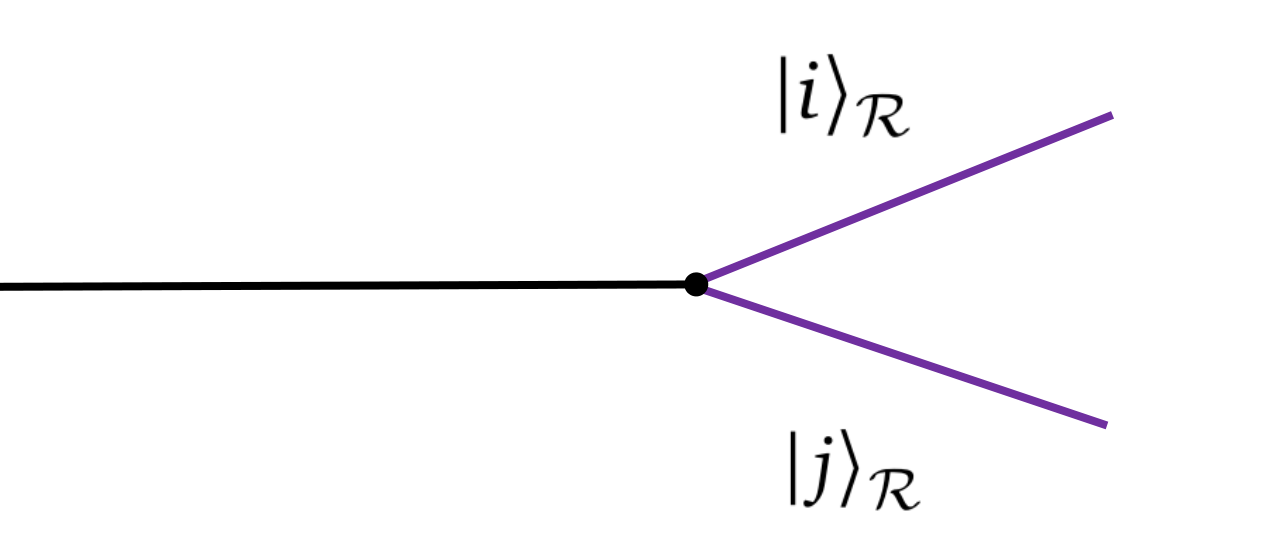}
	\caption{$\rho_{\mathcal{R}ij}$ in ordinary quantum systems represented by performing path integral with certain boundary conditions for the open edges at the region $\mathcal{R}$}
	\label{fig:reduced1}
\end{figure}

Firstly, let us review the computation of the reduced density matrix of $\mathcal{R}$ when the total system $I\cup B\cup R$ is in a pure state $\rho=\ket{\Psi}\bra{\Psi}$, where
\begin{align}
	\ket{\Psi}=\sum_{i,j,k}C_{ijk}\ket{i}_{I}\ket{j}_{B}\ket{k}_{\mathcal{R}},\qquad \sum_{i,j,k}C_{ijk}C_{ijk}^{*}=1\,.
\end{align}
Here $\{\ket{i}\},\{\ket{j}\},\{\ket{k}\}$ are the orthonormal bases of the Hilbert spaces $\mathcal{H}_{I}$, $\mathcal{H}_{B}$ and $\mathcal{H}_{R}$. In ordinary quantum systems, the degrees of freedom in different subsystems are independent in the sense that, the Hilbert space of the total system is factorized,
\begin{align}\label{Hfactorize}
	\mathcal{H}=\mathcal{H}_{I}\otimes\mathcal{H}_{B}\otimes\mathcal{H}_{R}\,.
\end{align}
The reduced density matrix of the subsystem $\mathcal{R}$ is then given by tracing out the degrees of freedom of the complement $I\cup B$ while setting boundary conditions for $\mathcal{R}$ with $\bra{k_{\mathcal{R}}}$ and $\ket{k'_{\mathcal{R}}}$,
\begin{align}
	(\rho_{\mathcal{R}})_{kk'}&=\sum_{i,j}\bra{k}_{\mathcal{R}}\bra{i}_{I}\bra{j}_{B}\rho\ket{j}_{B}\ket{i}_{I}\ket{k'}_{\mathcal{R}}=\sum_{i,j}\rho_{(ijk)(ijk')}
	\cr
	&=\sum_{i,j}C_{i,j,k}C_{i,j,k'}^{*}\,.
\end{align}
As we can see, any matrix element of the reduced density matrix $(\rho_{\mathcal{R}})_{kk'}$ is a summation of certain class of matrix elements of the density matrix $\rho$ of the total system, which is computed within the Hilbert space $\mathcal{H}$. This implies a summation of all possibilities outside $\mathcal{R}$ for a given set of boundary conditions on $\mathcal{R}$. For a local observer who can only measure the observables inside $\mathcal{R}$, the state of $\mathcal{R}$ is exactly given by the reduced density matrix $\rho_{\mathcal{R}}$. The entanglement entropy of $\mathcal{R}$ is then calculated by
\begin{align}
	S_{\mathcal{R}}\equiv S(\rho_{\mathcal{R}})=-\text{Tr}\left(\rho_{\mathcal{R}}\log\rho_{\mathcal{R}}\right)\,.
\end{align}
In quantum field theories, we use the path integral representation to compute the reduced density matrix \cite{Calabrese:2004eu,Calabrese:2009qy}. More explicitly, for scenarios with time reflection symmetry, $(\rho_{\mathcal{R}})_{ij}$ for $\mathcal{R}$ can be computed by cutting $\mathcal{R}$ open and setting different boundary conditions on the upper and lower edges, see Fig.\ref{fig:reduced1}. Then $\rho_{\mathcal{R}}^{n}$ is calculated by the replica trick via considering $n$ copies of the manifold and gluing them cyclically along the cuts present at $\mathcal{R}$. Upon taking the limit $n\to 1$ we get the entanglement entropy,
\begin{align}
	S(\rho_{\mathcal{R}})=S(\tilde{\rho}_{\mathcal{R}})=-\lim_{n\to 1}\partial_{n}\log \text{tr}(\rho_{\mathcal{R}}^{n}).
\end{align} 

The above paragraphs reviewed the standard way to compute the reduced density matrix and entanglement entropy in an ordinary quantum system. It is taken for granted that the degrees of freedom on the Cauchy slice are independent from each other and the Hilbert space factorizes following \eqref{Hfactorize}. 
Now we consider the self-encoded systems where such factorization no longer holds. 
{ We consider again a pure state of the system $I\cup B\cup\mathcal{R}$, but the system is highly constrained such that, for all the states in the reduced Hilbert space the state of the region $I$ is encoded in the state of $\mathcal{R}$ following a coding relation,
	\begin{align}\label{mapping}
		\ket{i}_{I}= f_{\mathcal{R}}\left(\ket{j}_{\mathcal{R}}\right)\,.
	\end{align}
	Compared with the generic coding relations \eqref{codingrelations}, here we have only one pair of $\mathcal{R}$ and $I$.}  As a mapping, we require 1) all the states in $\mathcal{H}_{\mathcal{R}}$ should be mapped to a unique state in $\mathcal{H}_{I}$, 2) all the states in $\mathcal{H}_{I}$ should have images in $\mathcal{H}_{\mathcal{R}}$.

The dimension of the reduced Hilbert space $\mathcal{H}_{red}$ decreases and $\mathcal{H}_{red}$ becomes a subspace of the original Hilbert space,
\begin{align}
	\text{dim}\mathcal{H}_{red}=\text{dim}\left(\mathcal{H}_{B}\otimes \mathcal{H}_{\mathcal{\mathcal{R}}}\right)\,,\qquad \mathcal{H}_{red}\subset \mathcal{H}_{I}\otimes\mathcal{H}_{B}\otimes\mathcal{H}_{R}\,.
\end{align}
One of the direct and crucial consequences of the self-encoding property \eqref{mapping} is that, some of the degrees of freedom become dependent on each other, hence  $\mathcal{H}_{red}$ is no longer factorizable following \eqref{Hfactorize}. In this case, the state of $I$ is determined by the state of $\mathcal{R}$ and hence $I$ does not add any independent degrees of freedom to the system.

To conduct an explicit calculation, let us consider the simplest configuration with two spins, where we can realize our previous statements. We denote the two spins as $I$ and $\mathcal{R}$ respectively, and denote the spin up (down) state as $\ket{0}$ ($\ket{1}$). 
At the beginning, we assume the system is in the pure state
\begin{align}\label{tspinstate}
	\rho=\ket{\Psi}\bra{\Psi}\,,\qquad \ket{\Psi}=\frac{\sqrt{2}}{2}\left(\ket{0_{I} 0_{\mathcal{R}}}+\ket{1_{I} 1_{\mathcal{R}}} \right)\,.
\end{align}	
Firstly, let us consider the familiar scenario where the spin $I$ is independent of $\mathcal{R}$. 
The Hilbert space $\mathcal{H}$ is four-dimensional with the following four orthonormal basis
\begin{align}\label{Hilberto}
	\mathcal{H}=\left\{\ket{0_{I}0_{\mathcal{R}}},\ket{0_{I}1_{\mathcal{R}}},\ket{1_{I}0_{\mathcal{R}}},\ket{1_{I}1_{\mathcal{R}}}\right\}\,.
\end{align}
The matrix elements of the reduced density matrix are given by,
\begin{align}
	\rho_{\mathcal{R}}{}_{00}=&\bra{0_{\mathcal{R}}0_{I}}\rho\ket{0_{I}0_{\mathcal{R}}}+\bra{0_{\mathcal{R}}1_{I}}\rho\ket{1_{I}0_{\mathcal{R}}}=\frac{1}{2}\,,
	\\\label{reducedmatrixelement1}
	\rho_{\mathcal{R}}{}_{01}=&\bra{0_{\mathcal{R}}0_{I}}\rho\ket{0_{I}1_{\mathcal{R}}}+\bra{0_{\mathcal{R}}1_{I}}\rho\ket{1_{I}1_{\mathcal{R}}}=0\,,
	\\
	\rho_{\mathcal{R}}{}_{10}=&\bra{1_{\mathcal{R}}0_{I}}\rho\ket{0_{I}0_{\mathcal{R}}}+\bra{1_{\mathcal{R}}1_{I}}\rho\ket{1_{I}0_{\mathcal{R}}}=0\,,
	\\	\rho_{\mathcal{R}}{}_{11}=&\bra{1_{\mathcal{R}}0_{I}}\rho\ket{0_{I}1_{\mathcal{R}}}+\bra{1_{\mathcal{R}}1_{I}}\rho\ket{1_{I}1_{\mathcal{R}}}=\frac{1}{2}\,.
\end{align}
The von Neumann entropy for $\rho_{\mathcal{R}}$ is then given by
\begin{align}
	S(\rho_{\mathcal{R}})=-\text{Tr} \left(\rho_{\mathcal{R}}\log\rho_{\mathcal{R}}\right)=\log 2\,,
\end{align} 
which indicates that the two spins are maximally entangled with each other.

\begin{figure}
	\centering
	\includegraphics[width=0.9\linewidth]{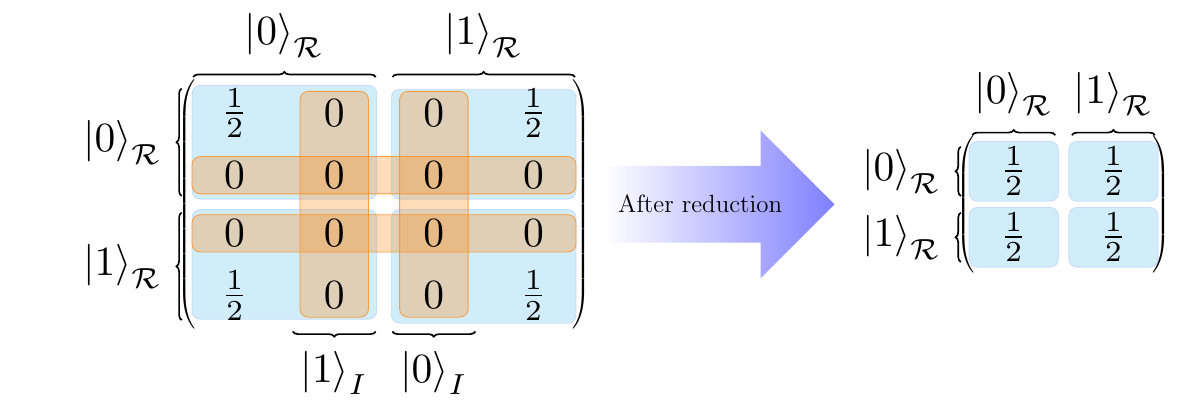}
	\caption{Illustration of the density matrice $\rho_{I\cup \mathcal{R}}$ for the two-spin system before(LHS) and after(RHS) the reduction of the Hilbert space.\label{fig:before_after_reduction_matrix}}
\end{figure}

Then, we consider the new configuration with constraints on the system such that the state of the spin $I$ is somehow totally determined by $\mathcal{R}$. { The constraints may be imposed by requiring that the $z$ component of the angular momentum should satisfy $J_{z}\neq 0$ all the time}\footnote{{ Perhaps it is easier to require $J_{z}=0$ such that the reduced Hilbert space is $\mathcal{H}_{red}=\left\{\ket{0_{I}1_{\mathcal{R}}},\ket{1_{I}0_{\mathcal{R}}}\right\}$, which implies the spin of the two spins should be opposite.}}. The constrains project out the states with $\left\{\ket{0_{I}1_{\mathcal{R}}},\ket{1_{I}0_{\mathcal{R}}}\right\}$ components, such that the four-dimensional Hilbert space \eqref{Hilberto} reduces to the following two-dimensional one
\begin{align}\label{Hilbertn}
	Reduced:\qquad\mathcal{H}_{red}=\left\{\ket{0_{I}0_{\mathcal{R}}},\ket{1_{I}1_{\mathcal{R}}}\right\}\,.
\end{align}
In this reduced Hilbert space, a coding relation emerges which indicates that the spin of $I$ must be the same as $\mathcal{R}$, i.e.
\begin{align}\label{coding1}
	\textit{Coding relation}:\qquad  \ket{0}_{I}=f_{\mathcal{R}}\left(\ket{0}_{\mathcal{R}}\right)\,,\qquad \ket{1}_{I}=f_{\mathcal{R}}\left(\ket{1}_{\mathcal{R}}\right)\,.
\end{align}	
Again, we consider the system to be in the state \eqref{tspinstate}, which is now a vector evolving in the reduced Hilbert space \eqref{Hilbertn} rather than the four-dimensional one \eqref{Hilberto}. Note that, the two states $\ket{0_{I}1_{\mathcal{R}}},\ket{1_{I}0_{\mathcal{R}}}$ are no longer basis of $\mathcal{H}_{I\cup \mathcal{R}}$, and the density matrix of the total system becomes $2\times 2$ dimensional, see  Fig.\ref{fig:before_after_reduction_matrix}.

Then how do we trace out the degrees of freedom for $I$ in the reduced Hilbert space? 
It turns out that, due to the constraints there is no room to perform the trace operation for $I$, { hence concept $\tilde{\rho}_{\mathcal{R}}$ is not a well-defined density matrix.} 
More explicitly, when we set boundary conditions for $\mathcal{R}$, we are fixing the state of $\mathcal{R}$. 
Since the state of $I$ is totally determined by $\mathcal{R}$ following the coding relation \eqref{coding1}, we simultaneously set boundary conditions on $I$. 
The reduced density matrix $\rho_{\mathcal{R}}$ is then calculated by,
\begin{align}
	\rho_{\mathcal{R}}{}_{00}=\bra{0_{\mathcal{R}}0_{I}}\rho\ket{0_{I}0_{\mathcal{R}}}=\frac{1}{2}\,,
	\\\label{reducedmatrixelementn1}
	\rho_{\mathcal{R}}{}_{01}=\bra{0_{I}0_{\mathcal{R}}}\rho\ket{1_{I}1_{\mathcal{R}}}=\frac{1}{2}\,,
	\\
	\rho_{\mathcal{R}}{}_{10}=\bra{1_{\mathcal{R}}1_{I}}\rho\ket{0_{I}0_{\mathcal{R}}}=\frac{1}{2}\,,
	\\
	\rho_{\mathcal{R}}{}_{11}=\bra{1_{I}1_{\mathcal{R}}}\rho\ket{1_{I}1_{\mathcal{R}}}=\frac{1}{2}\,,
\end{align}
and eventually we get (see Fig.\ref{fig:before_after_reduction_matrix})
\begin{align}
	\rho_{\mathcal{R}}=\rho_{I\cup \mathcal{R}}, \qquad\textit{and}\qquad S(\rho_{\mathcal{R}})=0\,.
\end{align}
One can further check that the von Neumann entropy for $\rho_{\mathcal{R}}$ is zero and hence $\rho_{\mathcal{R}}$ remains to be a pure state. This is expected, as we have mentioned that the additional spin $I$ does not add any independent degrees of freedom to the system. 
One may still be confused about the way we compute $\rho_{\mathcal{R}}$ and ask why we have not traced out the degrees of freedom of $I$ in \eqref{reducedmatrixelementn1} as we did in \eqref{reducedmatrixelement1}. The reason is that, the Hilbert space \eqref{Hilbertn} is reduced such that the terms in \eqref{reducedmatrixelement1} are no longer matrix elements of the density matrix $\rho$. If we insist to compute following \eqref{reducedmatrixelement1}, then states outside the reduced Hilbert space will be involved, which is not allowed by the constraints.

Though the two-spin system is extremely simple, we learn the following important lesson from it.
\begin{itemize}
	\item \textit{The reduced density matrix and relevant entropy quantities not only depend on the state of the system, but also depend on the Hilbert space where the state is embedded in. }
\end{itemize}
It is worth mentioning that, similar ideas has been used to clarify the ambiguity of the entanglement entropy in gauge theories (see \cite{Casini:2013rba,Ghosh:2015iwa} and especially \cite{Hung:2015fla}\footnote{At the final stage of this paper, professor Ling-yan Hung pointed out to us that, a similar discussion on the two-spin system is already given in \cite{Hung:2015fla} from the perspectives of what can be measured in an experiment.}). In gauge theories where the Hilbert space is usually redundantly labelled, the ambiguity of the entanglement entropy can naturally be understood as arising from the different choices of the Hilbert space where the state is embedded in.

\subsection{New island formula from Hilbert space reduction}
\begin{figure}[ht]
	\centering
	\includegraphics[scale=0.4]{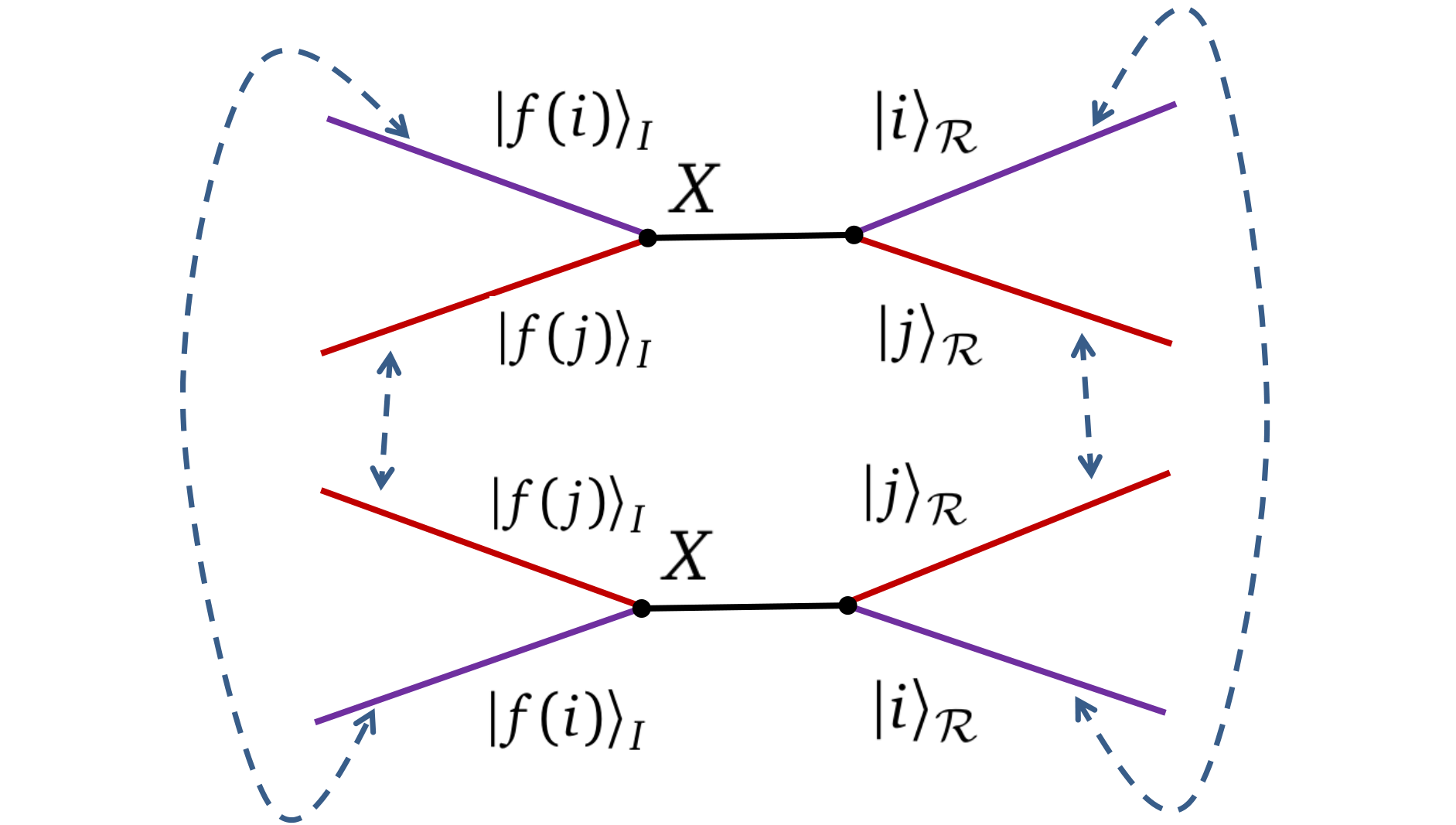}
	\caption{$\rho_{\mathcal{R}ij}$, $\rho_{\mathcal{R}ji}$ and $\text{Tr}\rho_{\mathcal{R}}^{2}$ in self encoding quantum systems with $I$ encoded in $\mathcal{R}$.}
	\label{fig:reduced2}
\end{figure}

Assuming that the self-encoding property can also be applied to QFT or gravity, we can generalize our discussion for the two-spin system a bit more to a QFT\footnote{In a QFT, placing a cutoff effectively discretizes the theory, enabling the application of finite-dimensional Hilbert space techniques.} or gravity in the path integral representation. Again we consider the simplest coding relation \eqref{mapping}. Unlike the two spin system, in general the total system also includes a region $B$ where the degrees of freedom are independent. To compute the reduced density matrix for $\mathcal{R}$ in the path integral description, we cut $\mathcal{R}$ open and set boundary conditions for the upper and lower edges at $\mathcal{R}$.  At the same time, we should simultaneously cut $I$ open and impose certain boundary conditions on the $I$ edges following the coding relation \eqref{mapping}.  Like the two-spin system case we discussed previously, the essential reason behind is that, we are doing computation in the reduced Hilbert space due to certain non-trivial constraints. See Fig.\ref{fig:reduced2} for a illustration of the reduced density matrix $\rho_{\mathcal{R}}{}_{ij}$.

Then we calculate the entanglement entropy via the replica trick, which glues the $n$ copies of the density matrix cyclically. 
Since certain boundary conditions are imposed on $I$ as we set boundary conditions on $\mathcal{R}$ due to the constraints, when the boundary conditions are settled such that we cyclically glue different copies of the system at $\mathcal{R}$, the corresponding boundary conditions at $I$ following the codes \eqref{mapping} also imply that we simultaneously glue different copies at $I$. 
In other words the cyclic gluing performed on $\mathcal{R}$ induces the cyclic gluing on $I$ (see Fig.\ref{fig:reduced2}). 
This results in an additional twist operator inserted at $X$, which is the boundary of $I$. 
In this scenario the region $I$ is nothing but the so-called ``\textit{entanglement island}'' in the literature. 
As in the two-spin system, if we insist to trace out the degrees of freedom on $I$, then the calculation will involve states outside the reduced Hilbert space $\mathcal{H}_{red}$, which is not allowed by the constraints. { 
In other words the notations $\tilde{\rho}_{\mathcal{R}}$ and $S(\tilde{\rho}_{\mathcal{R}})$ are ill-defined in the reduced Hilbert space. But the notation $\tilde{\rho}_{\mathcal{R}\cup I}$ can be well-defined as the degrees of freedom outside $\mathcal{R}\cup I$ is independent from $\mathcal{R}\cup I$. This is very similar to the case of $\tilde{\rho}_{\mathcal{R}\cup I}$ in gravitational theory, which can be taken as a density matrix when $I$ is the corresponding entanglement island of $\mathcal{R}$, see \eqref{rdmet}. }

Let us denote 
\begin{align}
	\tilde{S}_{\mathcal{A}}\equiv S(\tilde{\rho}_{\mathcal{A}})\,,
\end{align}
as the von Neumann entropy calculated by cyclically gluing only the region $\mathcal{A}$ in replica trick. In ordinary systems where the Hilbert space is not reduced, we have the trivial relation $S(\rho_{\mathcal{A}})\equiv S_{\mathcal{A}}=\tilde{S}_{\mathcal{R}}$. Nevertheless, in the self-encoded configurations we currently consider, this relation no longer holds. Based on the above discussions on the replica story, we arrive at the following crucial relation for self-encoded systems,
\begin{align}
	S_{\mathcal{R}}=\tilde{S}_{\mathcal{R}\cup I}\,.
\end{align}

Furthermore, if $I$ is settled on a gravitational background while $\mathcal{R}$ is settled in a non-gravitational bath, then when we apply the replica trick for $I\cup\mathcal{R}$ following \cite{Lewkowycz:2013nqa, Faulkner:2013ana}, we will receive additional gravitational contribution for $\tilde{S}_{\mathcal{R}\cup I}$, which is proportional to the area of the boundary $X$ of $I$, since $I$ is settled in a gravitational background. The non-gravitational part of $\tilde{S}_{\mathcal{R}\cup I}$ is just the bulk entanglement entropy $S_{bulk}({\tilde{\rho}_{\mathcal{R}\cup I}})$ which is calculated in a fixed curved background. Eventually we arrive at the following formula,
\begin{align}\label{islandformula2}
	Island~formula~\uppercase\expandafter{\romannumeral2}:\quad S_{\mathcal{R}}=\tilde{S}_{\mathcal{R}\cup I}=S_{bulk}({\tilde{\rho}_{\mathcal{R}\cup I}})+\frac{Area(X)}{4G}\,,\quad 	\ket{i}_{I}= f_{\mathcal{R}}\left(\ket{j}_{\mathcal{R}}\right)\,,
\end{align}
which exactly matches to the formula within the parenthesis in the \textit{island formula} $I$ in \eqref{islandformula1}. For multiple pairs of coding regions $\{I_a,\mathcal{R}_a\}$, the island formula $II$ is then generalized to a series of equations for any $\mathcal{R}_{a}$ with non-empty $I_{a}$,
\begin{align}\label{islandformula2g}
	Island~formula~\uppercase\expandafter{\romannumeral2}:\qquad	S_{\mathcal{R}_{a}}=S_{bulk}(\tilde{\rho}_{\mathcal{R}_{a}\cup\mathcal{I}_{a}})+\frac{Area(X_a)}{4G},\qquad 	\ket{i}_{I_{a}}= f_{\mathcal{R}_{a}}\left(\ket{j}_{\mathcal{R}_{a}}\right)\,.
\end{align} 
Note that, the formulas \eqref{islandformula2} and \eqref{islandformula2g} are based on the assumption that the system is self-encoded and we know the details of the coding relations \eqref{codingrelations}. If the assumed coding relation \eqref{codingrelations} is exactly the relation between the reservoir and its entanglement island, which solves the optimization problem of the island formula $I$ \eqref{islandformula1}, then the island formulas $I$ and $II$ are essentially the same. In the next section, we will give more discussion on their relationship.

\subsection{Requirements for the Hilbert space reductions}
The self-encoding constraints \eqref{mapping} is only one particular way to reduce the Hilbert space of the system. 
There are certainly other ways to reduce the Hilbert space, among which one will be introduced in later sections where the eliminated degrees of freedom are not localized in a definite spatial region. 
Nevertheless, not all reductions will essentially change the reduced density matrix and some of them are not even well defined. 
Here we present the following four requirements for the type of Hilbert space reductions which are interesting to us:
\begin{itemize}
	\item first of all, the state $\ket{\Psi}$ under consideration should remain in the reduced Hilbert space;
	
	\item secondly, when we impose boundary conditions for $\mathcal{R}$, we should have a square matrix block in $\rho$ from which the corresponding element of the reduced density matrix can be computed by tracing out the matrix block;
	
	\item thirdly, if we compute the reduced density matrix $\rho_{\mathcal{R}}$ in the reduced Hilbert space, we require that the dimension of $\rho_{\mathcal{R}}$ is still $\text{dim}\mathcal{H}_{\mathcal{R}}$, such that the local observer in $\mathcal{R}$ feel nothing wrong with the Hilbert space of $\mathcal{R}$. In other words, in the reduced Hilbert space of the total system the state of $\mathcal{R}$ is allowed to be any state in $\mathcal{H}_{R}$. 
	
\end{itemize}  

\begin{figure}
	\centering
	\includegraphics[width=0.9\linewidth]{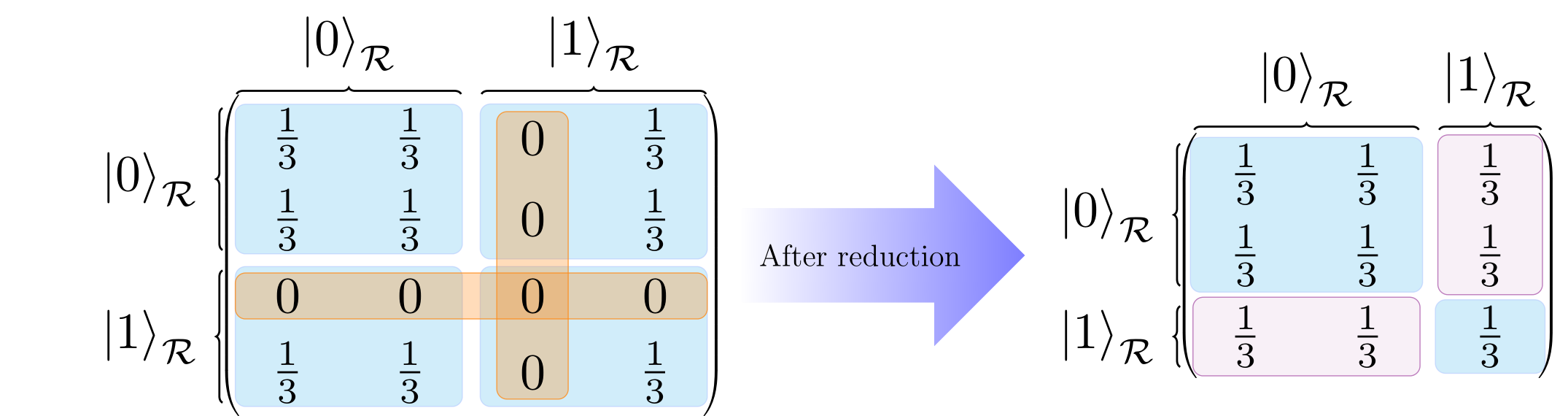}
	\caption{Illustration of the density matrices of $\ket{a}$ before(LHS) and after(RHS) the reduction of the Hilbert space.\label{fig:before_after_reduction_matrix1}}
\end{figure}

The second requirement implies that, after the reduction of the Hilbert space, the degrees of freedom for the complement $\bar{\mathcal{R}}$ should be preserved, no matter in which state the subsystem $\mathcal{R}$ is settled. For example, we can reduce the Hilbert space of the two-spins system to be
\begin{align}\label{reduced2}
	\mathcal{H}_{I\cup\mathcal{R}}=\{\ket{0_{I} 0_{\mathcal{R}}},\ket{1_{I} 0_{\mathcal{R}}},\ket{1_{I} 1_{\mathcal{R}}}\}\,,
\end{align}
in which $I$ is fixed to be the same as $\mathcal{R}$ when $\mathcal{R}$ is in the state $\ket{1_{\mathcal{R}}}$. 
On the other hand when $\mathcal{R}$ is in the state $\ket{0_{\mathcal{R}}}$, there is no constraint on the spin $I$. 
In other words the degrees of freedom of $I$ differ depending on the state of $\mathcal{R}$, hence our second requirement is not satisfied. 
This could be problematic, since when setting boundary conditions for $\mathcal{R}$ and computing the elements of $\rho_{\mathcal{R}}$, we will find that the matrix block is no longer a square matrix, which means that the degrees of freedom of $I$ cannot be described in the usual sense of a density matrix anymore. 
One should further study and define the physical meaning of density matrix blocks that are not square. 
Nevertheless, this is beyond the scope of this paper and we will naively consider such reductions to be un-physical. 
One necessary condition for the second requirement is that, the dimension of the reduced Hilbert space should be an integer times $\dim\mathcal{H}_{\mathcal{R}}$.
As an explicit example, we consider a vector $\ket{a}$ in the Hilbert space of the two-spins system before reduction
\begin{align}
	\ket{a}=\frac{1}{\sqrt{3}}\left(\ket{0_I0_\mathcal{R}}+\ket{1_I0_\mathcal{R}}+\ket{1_I1_\mathcal{R}}\right)\,,
\end{align}
and the corresponding density matrix in the unreduced Hilbert space can be worked out as depicted on the left panel of Fig.\ref{fig:before_after_reduction_matrix1}. 
After reducing the Hilbert space to $\mathcal{H}_{I\cup\mathcal{R}}$ in \eqref{reduced2} in which the degrees of freedom of $I$ depend on the state in $\mathcal{R}$, the density matrix can be shown to be given in the form of the right panel of Fig.\ref{fig:before_after_reduction_matrix1}. 
The matrix elements in the orange area all vanish since the state $\ket{a}$ does not contain any $\ket{0_I 1_\mathcal{R}}$ component, which is the constraint from our first requirement.  
The matrix blocks enclosed by the purple areas are not square which is different from density matrices before the reduction of Hilbert space. 
Thus, the reduced density matrix after the reduction of the Hilbert space is not well-defined in the usual way and this is exactly the point of our second requirement.  

\subsection{Islands beyond spatial regions}
Now we introduce a class of reductions where the reduced degrees of freedom in $\bar{\mathcal{R}}$ are not localized in a spatial subregion $I$. Let us use two sets of parameters $\{a_i\}$ and $\{b_j\}$ to denote the states in the Hilbert space $\mathcal{H}_{\bar{\mathcal{R}}}$ as follows
\begin{align}
	\ket{a_1,a_2\cdots a_n,b_1,b_2\cdots b_m}_{\bar{\mathcal{R}}}\in \ket{\mathcal{V}_{a}\otimes\mathcal{V}_{b}}=\mathcal{H}_{\bar{\mathcal{R}}}\,.
\end{align}
and a generic state for the entire system can be expressed by
\begin{align}
	\ket{\Psi}=\sum_{a_1,\cdots a_n,b_1,\cdots b_m, i}C_{a_1,\cdots a_n,b_1,\cdots b_m, i}\ket{a_1,\cdots a_n,b_{1},\cdots b_{m}}_{\bar{\mathcal{R}}}\ket{i}_{\mathcal{R}}\,.
\end{align}
Here, for example, $\mathcal{V}_{a}$ represents a vector space in which a generic vector can be specified by the set of parameters $\{a_i\}$. 
It is not the Hilbert space of any region, since $a$ is not a subregion of $\bar{\mathcal{R}}$. 

Now we reduce the Hilbert space following certain constraints. 
We assume that the state of $\mathcal{R}$ determines the parameters $\{a_i\}$ in the following way
\begin{align}\label{reductionnonlocal}
	\ket{a_1,a_2\cdots a_n,b_1,b_2\cdots b_m}_{\bar{\mathcal{R}}}= \ket{\vec{a}(\ket{i}_{\mathcal{R}}),b_{1},b_{2}\cdots b_{m}}_{\bar{\mathcal{R}}}\,,
\end{align}
where $\vec{a}$ is a vector in the space $\mathcal{V}_{a}$. 
This means that for any state $\ket{\Psi}$ in the reduced Hilbert space, if the state $\ket{i}_{\mathcal{R}}$ of the subregion $\mathcal{R}$ is given, then the corresponding state of $\bar{\mathcal{R}}$ is partially determined with the parameters in the subspace $\mathcal{V}_{a}$ fixed to be $\vec{a}(\ket{i}_{\mathcal{R}})$. 
Hence the dimension of the Hilbert space reduces to be $\text{dim}(\mathcal{V}_{b})\times \text{dim}(\mathcal{H}_{\mathcal{R}})$. 
In the reduced Hilbert space, a generic state $\ket{\Psi}$ can be expressed by
\begin{align}
	\ket{\Psi}=\sum_{b_1,b_2\cdots b_m, i}C_{b_1,b_2\cdots b_m, i}\ket{\vec{a}(\ket{i}_{\mathcal{R}}),b_{1},b_{2}\cdots b_{m}}_{\bar{\mathcal{R}}}\ket{i}_{\mathcal{R}}\,.
\end{align}
Note that, in the reduced Hilbert space the independent degrees of freedom are confined in the subspace $\mathcal{V}_{b}$, which are usually parameters characterizing the state of $\bar{\mathcal{R}}$, rather than a subset $I$ inside $\bar{\mathcal{R}}$. 
Then the reduced density matrix is calculated by
\begin{align}
	(\rho_{\mathcal{R}})_{ij}=\sum_{b_1,b_2\cdots b_m}C_{b_1\cdots b_m, i}C^{*}_{b_1\cdots b_m, j}\ket{\vec{a}(\ket{i}_{\mathcal{R}}),b_{1},\cdots b_{m}}_{\bar{\mathcal{R}}}\bra{\vec{a}(\ket{j}_{\mathcal{R}}),b_{1},\cdots b_{m}}_{\bar{\mathcal{R}}}\,,
\end{align}
where we have only traced the independent degrees of freedom in $\mathcal{V}_{b}$.

In this type of reductions \eqref{reductionnonlocal} the reduced degrees of freedom in $\mathcal{V}_{a}$ are not required to be mapped to the degrees of freedom localized in any spatial region $I$ inside $\bar{\mathcal{R}}$, hence there could be no spatial region $I$ that plays the role of an island. 
Rather, the island is a sub-space in the parameter space which characterize the Hilbert space.

\section{Relation to the \textit{island formula} $I$ in gravitational theories}\label{sec3}

The main motivation for our mechanism comes from our gradually deepened understanding about the quantum information aspects of gravity. Firstly, there is a basic puzzle of gravity that accompany the Bekenstein-Hawking formula, which is the fact that the number of degrees of freedom in the black hole interior is divergent if we take the effective field theory of gravity as a quantum field theory (with gravitational fluctuations), while the number indicated by the Bekenstein-Hawking entropy
\begin{align}\label{BHentropy}
	S_{BH}=\frac{Area (\partial A)}{4G}\,,
\end{align}	
is finite. If we introduce a UV cut-off to regularize the entropy, it is even more puzzling as we note that the Bekenstein-Hawking entropy is independent from the UV cutoff. This puzzle is also a long-standing problem of interpreting the black hole entropy as the entanglement entropy between the black hole interior and outside \cite{Dvali:2008jb,Solodukhin:2011gn}. Then a possible solution to this puzzle in following.
\begin{itemize}
	\item \textit{The Hilbert space of the ``quantum'' gravity is a reduced space of the Hilbert space of the corresponding local quantum field theory with gravitational fluctuations \footnote{{  The degree of the Hilbert space reduction directly affects the number of degrees of freedom in quantum gravity, which is further related to the cutoff scale of the theory. Regarding the cutoff scale of quantum gravity, there have been discussions on the relation between Newton's constant, the number of species of particles, and the so-called gravity cutoff $\Lambda$, see \cite{Hawking:2000da,Dvali:2007hz,Dvali:2008jb} for original papers. A more tempting story is that, the bulk entanglement entropy plus the Bekenstein-Hawking entropy gives the renormalized entropy $\frac{Area(A)}{4G}$ with a renormalized Newton constant. Nevertheless, such renormalization is only successful for minimally coupled matter fields (see \cite{Solodukhin:2011gn} for details), if gravity is coupled to non-minimally coupled matter fields, like gauge bosons, this story is not well understood. }}.}
\end{itemize}

Then, it is very interesting to ask whether the reduction of Hilbert space induces self-encoding property of gravity. According to our discussion in the previous section, it seems that, the self-encoding property is the only plausible interpretation for the emergence of entanglement islands from the quantum information perspective. When entanglement entropy is calculated by the island formula $I$ in gravitational theories, it implies that the gravitational theory is self-encoded.

In fact the conjecture that gravitational theories could be self-encoded turns out to be one of the most important implication from the developments in the quantum information aspects of quantum gravity in recent two decades. It was first conjectured in several independent papers \cite{Bousso:2012as,Verlinde:2012cy,Papadodimas:2012aq,Susskind:2013tg,Maldacena:2013xja} that the state of the black hole interior is mapped to the state of the Hawking radiation after the Page time, hence they are not independent degrees of freedom. This solves the Mathur/AMPS puzzle \cite{Mathur:2009hf,Almheiri:2012rt}, as the monogamy of entanglement only applies to independent degrees of freedom. This is nothing but the self encoding-property we have introduced.

This idea was sharpen in the double holography set-up \cite{Almheiri:2019hni}. Given a region $\mathcal{R}_{a}$ in the non-gravitational bath the corresponding island region $I_{a}$ could be included as part of the entanglement wedge of $\mathcal{R}_{a}$. According to the ``entanglement wedge reconstruction'' program  \cite{Bousso:2012mh,Czech:2012bh,Wall:2012uf,Headrick:2014cta,Almheiri:2014lwa,Jafferis:2015del,Dong:2016eik,Harlow:2016vwg,Faulkner:2017vdd}, the states and operators in the entanglement wedge (which includes $I_{a}$) of $\mathcal{R}_{a}$ can be reconstructed from the information in $\mathcal{R}_{a}$. In the more generic configurations without assuming holography, the pairs of $\{\mathcal{R}_{a},I_{a}\}$ regions are both subregions in gravitational theories, rather than on different sides of holography. The existence of reconstruction can also be justified following the same line of argument for entanglement wedge reconstruction. More importantly, an explicit reconstruction of the island region $I_{a}$ can be derived via the Petz map \cite{Petz:1986tvy,Petz:1988usv,Cotler:2017erl,Chen:2019gbt} in some simple models of quantum gravity \cite{Penington:2019kki}, and this reconstruction in principle can be generalized to more generic gravitational theories. 

Independently, in another series of papers \cite{Laddha:2020kvp,Chowdhury:2020hse,Chowdhury:2021nxw} it was proposed that, in gravitational theories all the information available on a Cauchy slice is also available near its boundary, which means the physics in any bulk regions away from the boundary can be reconstructed from the information in the near boundary region. 

\begin{itemize}
	\item \textit{In summary these important works imply that, gravitational theories are self-encoded in the sense of \eqref{codingrelations}. More explicitly, the regions $\mathcal{R}_{a}$ and $I_{a}$ in the island formula $I$ satisfy a coding relation \eqref{codingrelations}, where the region correspondence $\{\mathcal{R}_{a},I_{a}\}$ is determined by the optimization procedure, and the state correspondence $\{\ket{j}_{\mathcal{R}_{a}},\ket{i}_{I_{a}}\}$ is implied by the reconstruction.}
\end{itemize}

The above review implies that like the island formula $II$, the island formula $I$ is also an result of the self-encoding property. Nevertheless, an obvious difference between the island formula $I$ \eqref{islandformula1} and $II$ \eqref{islandformula2g} is that, in \eqref{islandformula1} there is an optimization procedure which is missing in \eqref{islandformula2g}. Given a region $\mathcal{R}_{a}$, the optimization procedure determines its island region $I_{a}$, which is indeed the first level of the coding relation, which means the island formula $I$ contains more information than the island formula $II$, due to the special structure of spacetime wormholes in gravitational theories. Also, the Hilbert space reduction in gravity is not induced by any external constraints, rather it also seems to be a result of the wormhole structures which is an intrinsic property of gravity. {  All in all, the Hilbert spaces of gravitational theories reduce in a special type of ways for some unknown intrinsic reason, which gives specific coding relations and the mapping between $\mathcal{R}_a$ and $I_a$ can not be randomly chosen. In other words, when gravitation is introduced in the system, \eqref{islandformula2g} only makes sense when $I_a$ is exactly the entanglement island of $\mathcal{R}_a$.}

{ We give some more comments on describing the black hole evaporation process in  terms of Hilbert space reduction. From the formation to the Page time, the Hawking radiation $\mathcal{R}$ admits no islands, which indicates all the degrees of freedom outside $\mathcal{R}$ are (totally) independent from $\mathcal{R}$. After the page time, $\mathcal{R}$ admits a island $I$ in the black hole interior. This is a phase transition from the fact that $\mathcal{R}$ determines (almost) nothing to $\mathcal{R}$ totally determines $I$. We propose that this sudden change is a result of the large central charge limit. }

\section{A simulation of Hilbert space reduction via Weyl transformation}\label{sec:4-examples}
In the previous section, we have reviewed the clues demonstrating that gravitational theory has a reduced Hilbert space featured by finite cutoff, and show the possibility that this reduction induces the self-encoding property of gravity and the emergence of entanglement islands. Inspired by these clues, as well as the double holography \cite{Almheiri:2019hni} set-up, here we propose a prescription to simulate a configuration where island phases are extensively studied, i.e. the AdS/BCFT correspondence \cite{Takayanagi:2011zk}. We start from the vacuum state of the holographic CFT$_2$, then we introduce the finite cutoff to the theory by imposing a special Weyl transformation to half of the system $x<0$. { In order to apply the island formula $I$ on this Weyl transformed system, we assume that the $x<0$ of the theory is coupled to AdS$_2$ gravity. Furthermore, if we assume that the scalar field that characterizes the Weyl transformation is dynamical, the coupled AdS$_2$ gravity may be induced by the dynamics of the scalar field.} We will discuss the entanglement structure, application of the island formula $I$ and the emergence of entanglement islands in this configuration, and find that the main features of the AdS/BCFT configuration are perfectly captured by the Weyl transformed CFT. 

Note that, a similar prescription was earlier proposed independently in \cite{Suzuki:2022xwv} with a different motivation, as well as some technical differences from our prescription.

\subsection{Finite cutoff from Weyl transformation in CFT$_2$}\label{subsection41}
Let us start from the vacuum state of a holographic CFT$_2$ on a Euclidean flat space\footnote{Here the overall factor $\frac{1}{\delta^2}$ is inspired by AdS/CFT and \cref{flatmetric} acts as the boundary metric corresponding to the dual AdS$_3$ geometry,
	\begin{align}\label{Poincare3}
		ds^2=\frac{\ell^2}{z^2}(-dt^2+dx^2+dz^2)\,,
	\end{align}
	with the cutoff settled at $z=\delta$.}
\begin{align}\label{flatmetric}
	ds^2=\frac{1}{\delta^2}\left(d\tau^2+dx^2\right)\,.
\end{align}
Here $\delta$ is an infinitesimal constant representing the UV cutoff of the boundary CFT. The theory is invariant under the Weyl transformation of the metric 
\begin{align}
	ds^2=e^{2\varphi(x)}\left(\frac{d\tau^2+dx^2}{\delta^2}\right)\,.\label{Weyl}
\end{align}
This effectively changes the cutoff scale in the following way
\begin{align}
	\delta\Rightarrow e^{-\varphi (x)}\delta\,.
\end{align}
The entanglement entropy of a generic interval $\mathcal{R}=[a,b]$ in the CFT after the Weyl transformation picks up additional contributions from the scalar field ${\varphi}(x)$ as follows \cite{Caputa:2017urj,Caputa:2018xuf,Camargo:2022mme}
\begin{align}\label{entropyweyl}
	S_{\mathcal{R}}=\frac{c}{3}\log\left(\frac{b-a}{\delta}\right)+\frac{c}{6}{\varphi}(a)+\frac{c}{6}{\varphi}(b)\,.
\end{align}
This formula can be achieved by performing the Weyl transformation on the two-point function of the twist operators.

For our purpose to study entanglement islands, we perform a specific Weyl transformation on a holographic CFT$_2$ that corresponds to Poincar\'e AdS$_3$, such that the metric becomes AdS$_2$ in the $x<0$ region and remains flat in the $x>0$ region. Such a Weyl transformation can be easily found to be\footnote{Note that, at $x=0$ the scalar field \eqref{varphi1} is not smooth or even continuous. The entropy formula \eqref{entropyweyl} only depends on the scalar field at the endpoints, so we think this is not a problem as long as we do not talk about the intervals ending at $x=0$. We can also redefine the scalar field in the neighborhood of $x=0$ to retain smoothness there.}
\begin{align}\label{varphi1}
	\varphi(x)=
	\begin{cases}
		0\, &, \text{if} \quad x>0\,,\\ \\
		-\log\left(\frac{2|x|}{\delta}\right)+\kappa\,.\, &, \text{if} \quad x<0\,,
	\end{cases} 
\end{align}
where $\kappa$ is an undetermined constant. The corresponding metric at $x<0$ after the Weyl transformation becomes
\begin{align}
	ds^2=\frac{e^{2 \kappa }}{4}\left(\frac{d\tau^2+dx^2}{x^2}\right)\,,\qquad x<0\,.
\end{align}
As expected, the above metric is AdS$_2$ up to an overall coefficient (the length scale is given by $e^{\kappa}/2$). Note that, in order to get a AdS$_2$ metric independent of $\delta$, we need to choose a scalar field depending on $\delta$ which goes to infinity when $\delta\to 0$. The specific choice \eqref{varphi1} for the scalar field is made to simulate the entanglement structure for the AdS/BCFT correspondence, where the \textit{effective theory} description can be taken as an AdS$_2$ gravity coupled to a CFT$_2$ bath. Also the cutoff scale in the $x<0$ region is no longer infinitesimal, rather the cutoff scale is bounded from below and depends on the position $x$.

Before we go ahead, we give a physical interpretation for the Weyl transformation, as well as the entropy formula \eqref{entropyweyl}. Before the Weyl transformation the UV cutoff of the CFT is a uniform infinitesimal constant $\delta$. The Weyl transformation is indeed a position-dependent scale transformation that changes the cutoff scale from $\delta$ to some finite scale. Such a non-uniform cutoff would definitely dramatically reduces the Hilbert space as the UV degrees of freedom, as well as the small scale entanglement structure, are erased by the Weyl transformation. After the Weyl transformation, the formula \eqref{entropyweyl} tells us that the entanglement entropy is just the original one subtracting two constants\footnote{We require that the two endpoints are not close enough to give a negative entanglement entropy following \eqref{entropyweyl}.} which are totally determined by the scalar field at the two endpoints. More importantly the subjected constant is independent from the position of the other endpoint. Hence we conclude that given a particular point $x$, the Weyl transformation at $x$ effectively excludes all the small distance entanglement across $x$ below the cutoff scale, which is captured by the constant $\frac{c}{6}|\varphi (x)|$, while keeping the long distance entanglement structure above the cutoff scale unaffected.

\subsection{Weyl transformed CFT vs AdS/BCFT}
\begin{figure}[ht]
	\centering
	\includegraphics[scale=0.4]{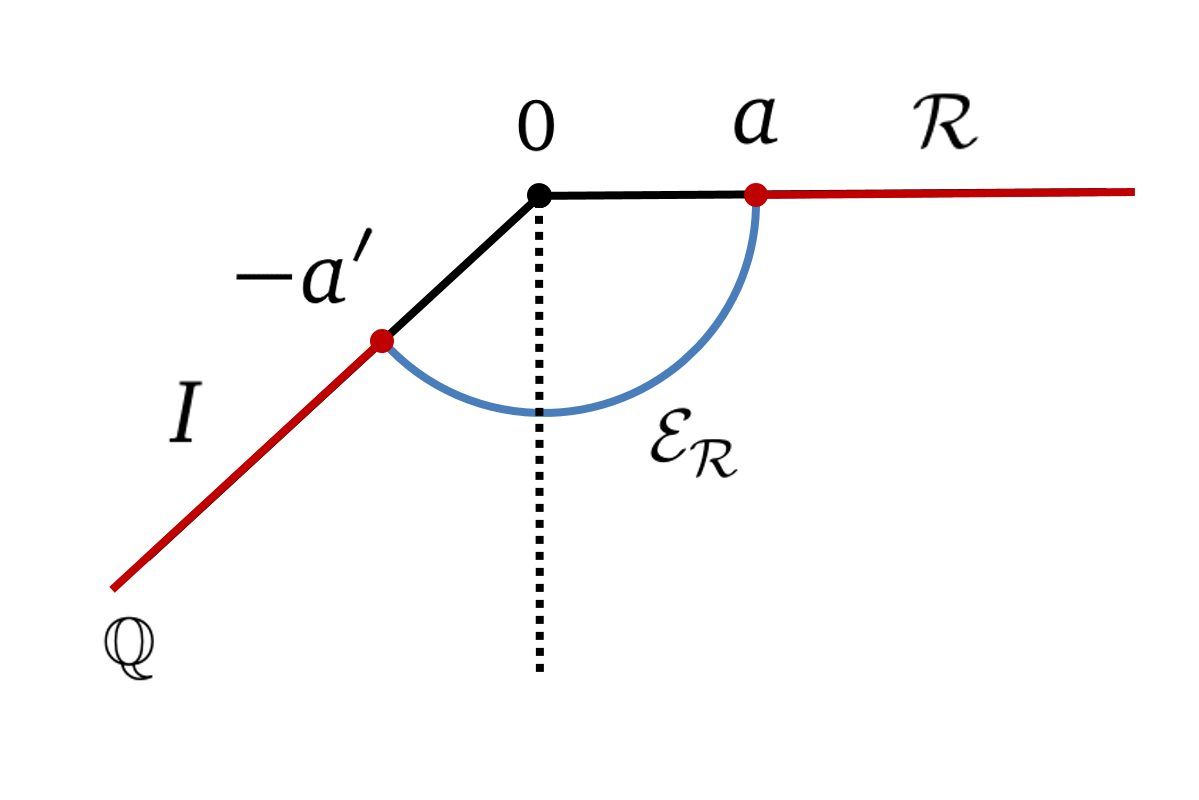}\quad \includegraphics[scale=0.4]{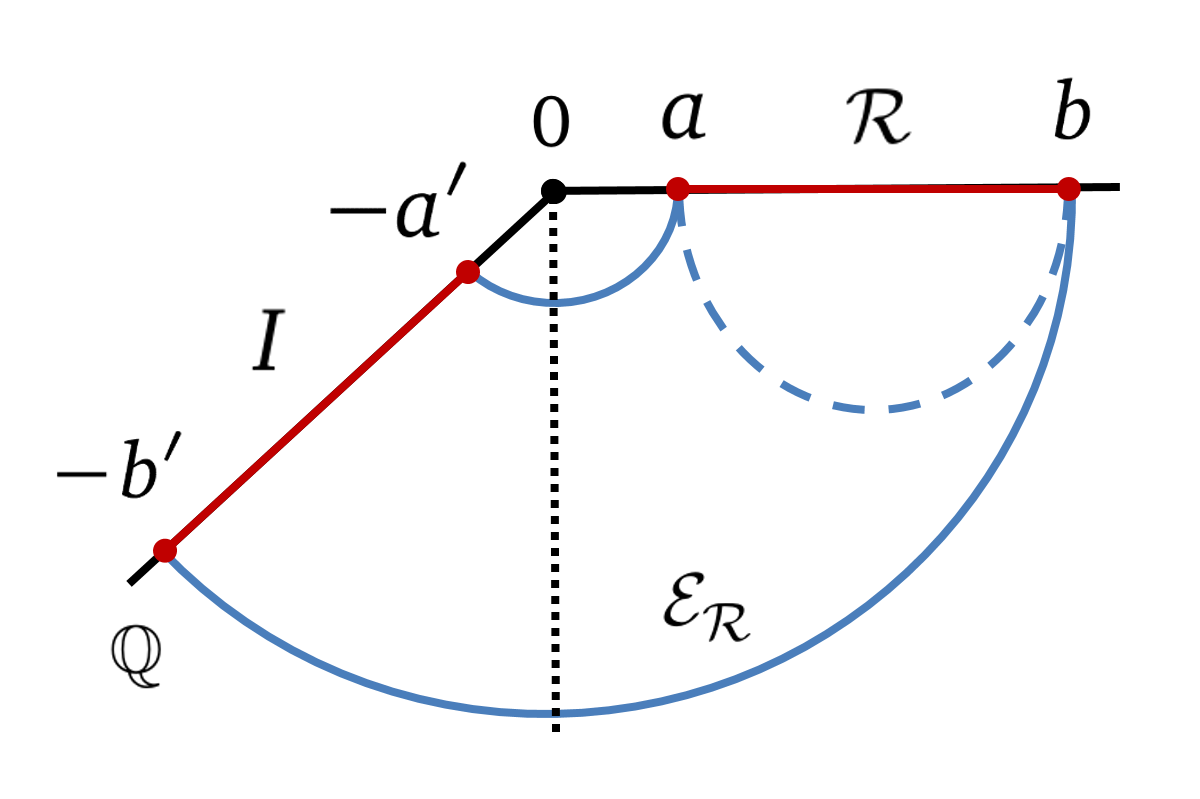}
	\caption{Holographic entanglement entropy in the AdS/BCFT setup: (left) For an interval $\mathcal{R}=[a,\infty]$ in the BCFT, the RT surface has the disconnected topology and lands on the EoW brane at $a'=a$. (right) For $\mathcal{R}=[a,b]$, the connected(disconnected) RT surfaces are shown by the blue dashed(solid) curves. The position where $\mathcal{E}_{\mathcal{R}}$ anchors on $\mathbb{Q}$ is determined by extremizing the length of $\mathcal{E}_{\mathcal{R}}$, which gives $a'=a$ and $b'=b$. The dashed vertical line denote the $\rho=0$ slice.}
	\label{fig:EE}
\end{figure}

The AdS/BCFT \cite{Takayanagi:2011zk} correspondence is a widely used setup where entanglement islands emerge. The basic statement is that, the holographic CFT$_2$ with a boundary correspond to the AdS$_3$ bulk with a end of world (EoW) brane which extends in the bulk and anchors on the boundary of the CFT. In the AdS$_3$ bulk, the EoW brane satisfies Neumann boundary condition and its position is determined by its tension. In this setup, it is more convenient to use another set of coordinates $(t,\rho,y)$ to describe the AdS$_3$ bulk geometry,
\begin{align}\label{x-y-coordinates}
	x=y \tanh\rho~~,~~z=-y \text{sech}\,\rho\,.
\end{align}
The bulk metric in these coordinates is given by the standard Poincar\'e slicing, as follows
\begin{align}
	ds^2&=d\rho^2+\cosh^2\rho\frac{-dt^2+dy^2}{y^2}\notag=\frac{-dt^2+dx^2+dz^2}{z^2}\,,
\end{align}
where the AdS$_3$ radius is set to be unity. In the Poincar\'e slicing\footnote{A convenient choice for a polar coordinate is $\theta=\text{arccos}\left(\text{sech}\,\rho\right)$, which determines the angular position of the brane from the vertical.} described by the $(t,\rho,y)$ coordinate chart the EoW brane is situated at a constant $\rho=\rho_0$ slice \cite{Fujita:2011fp}, where $\rho_{0}$ is determined by the tension $T$ of the EoW brane,
\begin{align}
	\rho_0= \text{arctanh} T\,.
\end{align}
It is easy to see that the metric on the EoW brane is exactly AdS$_2$ up to an overall constant.
The key property in the AdS/BCFT setup is that, the RT surface $\mathcal{E}_{\mathcal{R}}$ of a boundary region $\mathcal{R}$ is also allowed to be anchored on the EoW brane $\mathbb{Q}$ \cite{Takayanagi:2011zk,Fujita:2011fp} (see Fig.\ref{fig:EE}). This was confirmed in \cite{Sully:2020pza} via a direct computation of the correlation functions of twist operators in BCFTs with large central charge. Hence, new configurations for the RT surfaces that anchors on $\mathbb{Q}$ arise in AdS/BCFT. The island formula in double holography setups is reproduced by considering these new configurations of the RT surfaces when applying the RT formula to calculate the holographic entanglement entropy.

For example, the entanglement entropy of the region $\mathcal{R}:=[a,\infty)$ may be computed through the length of the RT curve $\mathcal{E}_{\mathcal{R}}$ homologous to $\mathcal{R}$ and anchored on the EoW brane. After determining the location on the brane by extremizing the length of $\mathcal{E}_{\mathcal{R}}$ (see the left panel in Fig.\ref{fig:EE}), the holographic entanglement entropy is given by
\begin{align}
	S_{\mathcal{R}}=\frac{Area(\mathcal{E}_{\mathcal{R}})}{4 G}=\frac{c}{6}\log\left(\frac{2a}{\delta}\right)+\frac{c}{6}\rho_0\,.\label{SR1}
\end{align}
Note that the area term $\frac{Area(\mathcal{E}_{\mathcal{R}})}{4 G}$ captures the bulk entanglement entropy term in the \textit{Island formula \uppercase\expandafter{\romannumeral1}}, while the area term $\frac{Area X}{4G}$ is ignored by taking the gravity theory settled on the EoW brane to be the induced gravity by partial reduction on the $\rho_0>\rho>0$ region. This choice is necessary to compare with the configuration of the holographic Weyl transformed CFT$_2$ later. For the choice $\mathcal{R}:=[a,b]$ there are two possible saddles for the RT surface, one is a connected geodesic that anchored on the two endpoints of $\mathcal{R}$, while the other consists of two disconnected geodesics which also anchor on the EoW brane $\mathbb{Q}$ (see the right panel in Fig.\ref{fig:EE}). The holographic entanglement entropy is simply given by
\begin{align}\label{varphi2}
	S_{\mathcal{R}}=
	\begin{cases}
		\frac{c}{3}\log\left(\frac{b-a}{\delta}\right)\, &,\, \text{if} \quad a>b \left(1-2 \sqrt{e^{2 \rho_0}+e^{4 \rho_0}}+2 e^{2 \rho_0}\right)\,,\\ \\
		\frac{c}{6}\log\left(\frac{2a}{\delta}\right)+\frac{c}{6}\log\left(\frac{2b}{\delta}\right)+\frac{c}{3}\rho_0\,.\, &,\, \text{if} \quad 0<a<b \left(1-2 \sqrt{e^{2 \rho_0}+e^{4 \rho_0}}+2 e^{2 \rho_0}\right)\,.
	\end{cases} 
\end{align}
In both left and right panels in Fig. \ref{fig:EE}, the portion of the brane $\mathbb{Q}$ (marked red) lying in the entanglement wedge of $\mathcal{R}$ are be interpreted as the island region from a doubly holographic point of view.

\begin{figure}[ht]
	\centering
	\includegraphics[scale=0.4]{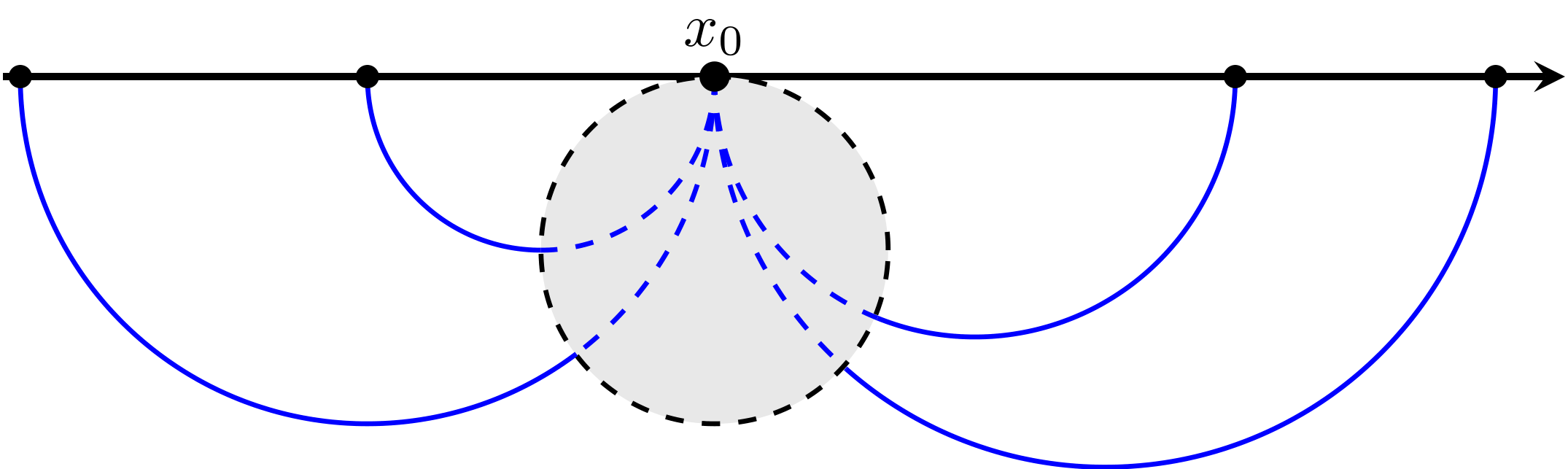}
	\caption{Cutoff sphere at $x=x_0$ in the bulk dual geometry for a Weyl transformed CFT. For a minimal surface anchored at $x_0$, the portion inside the cutoff sphere is excluded.}
	\label{fig:cutoffsphere}
\end{figure}

Now we return to the Weyl transformed CFT and compare with the version of island formula in the AdS/BCFT or doubly holographic setup. The physical interpretation for Weyl transformation is more intuitive when the CFT is holographic, hence the entanglement structure has a geometric interpretation. Before we perform the Weyl transformation, the vacuum state of the CFT is dual to Poincar\'e AdS$_3$ \eqref{Poincare3}. And according to the RT formula the entanglement entropy for any interval is given by the length of the minimal bulk geodesic homologous to this interval. 
The integral computing the length of the RT surface represents the collection of the entanglement at all the scales \cite{Swingle:2010jz}. In this context, the Weyl transformation adjusts the cutoff scale by adjusting the position of the cutoff point on the RT curve, where we stop integrating the length of the RT curve. According to the formula \eqref{entropyweyl}, for any RT surface anchored at the site $x_0$ on the boundary, we need to push the cutoff point on the RT surface from $z=\delta$ to certain position in the bulk, such that the length of the RT surface is reduced by certain constant $|{\varphi}(x_0)|$. In other words the cutoff points for all the RT curves anchored at $(\delta,x_0)$ form a sphere in the bulk.
Interestingly, for static configurations, the cutoff sphere in the AdS$_3$ background is a circle in flat background,
\begin{align}
	\left(x-x_0\right){}^2+\left(z-|x_0| e^{-\kappa }\right)^2=|x_0|^2 e^{-2\kappa } \,,
\end{align}
with the center being $(|x_0| e^{-\kappa },x_0)$ and the radius $r=|x_0| e^{-\kappa }$. One may consult Appendix \ref{appendix} for the derivation. The formula \eqref{entropyweyl} then can be understand as follows: when we integrate the length of the RT surface, we only integrate up to the cutoff sphere (see Fig.\ref{fig:cutoffsphere}).

\begin{figure}[ht]
	\centering
	\includegraphics[scale=0.45]{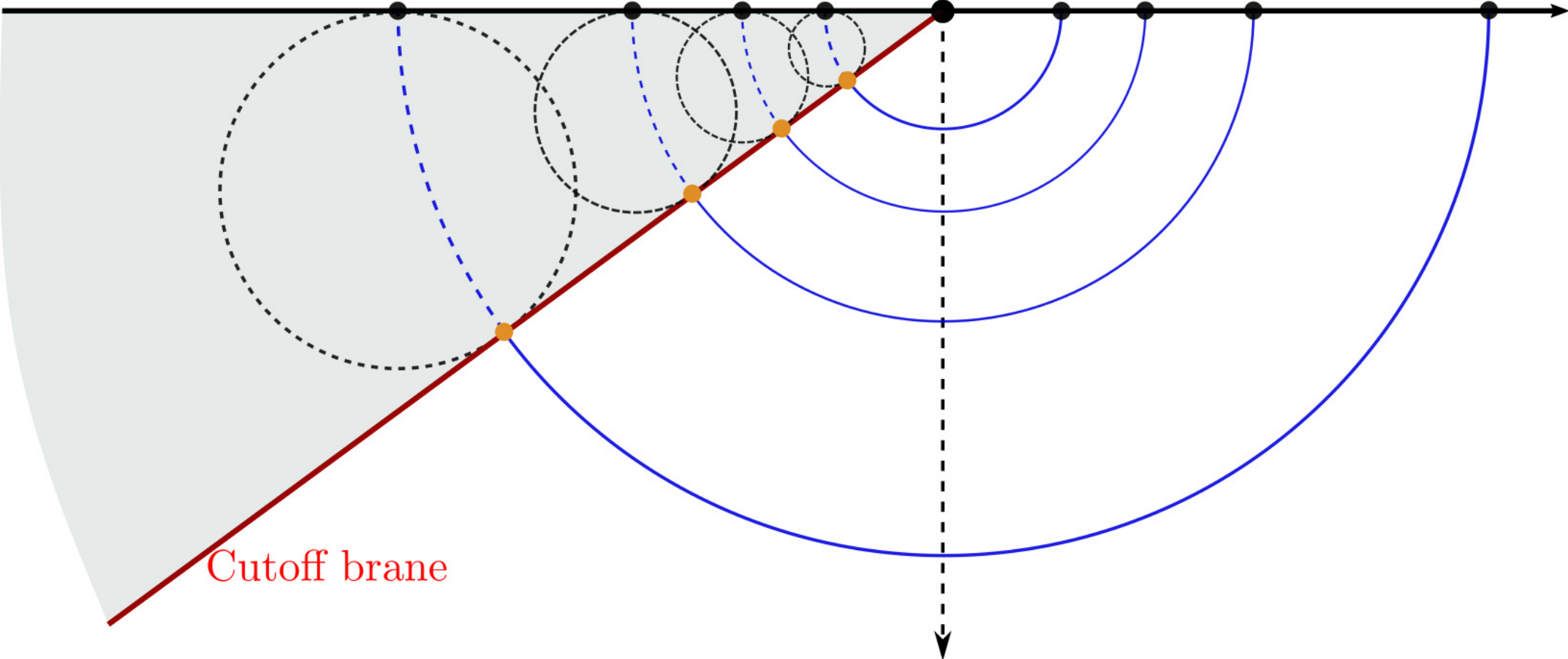}
	\caption{The cutoff brane as the common tangent line of all the cutoff spheres, and its comparison with the EoW brane in the AdS/BCFT scenario.}
	\label{fig:weylcftbcft}
\end{figure}

The Weyl transformation \eqref{varphi1} then adjusts the cutoff scale in the region $x<0$. In some sense, this pushes the cutoff point from $z=\delta$ into the bulk on the cutoff spheres. Here we define the cutoff brane:
\begin{itemize}
	\item \textit{Cutoff Brane: the common tangent line of all the cutoff spheres. For the specific Weyl transformation characterized by \eqref{varphi1}, the cutoff brane is depicted by (see appendix \ref{appendix} for details)
		\begin{align}
			\rho=\kappa\,.
	\end{align}}
\end{itemize}
This reminds us of the EoW brane in the AdS/BCFT setup, which is also settled at a constant $\rho$ slice. When we calculate the holographic entanglement entropy for some boundary region $\mathcal{R}$ at $x>0$, we also consider all the possible configurations where the homologous surfaces anchor on the cutoff brane, then the RT surface is the one with the minimal length\footnote{The fact that the RT surface anchors on the cutoff brane can also be understood as minimizing the length of all homologous surfaces that anchors on all possible cutoff spheres.}. Then the cutoff brane indeed plays the same role as the EoW brane $\mathbb{Q}$ in AdS/BCFT \footnote{We may also consider the case of an EoW brane with an intrinsic Einstein-Hilbert term in an AdS$_2$ background, similar to the Dvali-Gabadadze-Porrati (DGP) gravity \cite{Dvali:2000hr}. However, as discussed in \cite{Chen:2020uac, Chen:2020hmv}, one can fine tune the brane tension to cancel the curvature contributions from the DGP term and hence keep the position of the brane fixed. In the cut-off brane picture, adding a DGP like term is therefore tantamount to a convenient choice of the Weyl factor $\varphi(x)$ corresponding to the same generalized entanglement entropy in the two perspectives.}, see Fig. \ref{fig:weylcftbcft}. When we set $\kappa=\rho_0$, then the holographic calculation for $S_{\mathcal{R}}$ is exactly the same as the calculation in AdS/BCFT, which we have just reviewed.

On the other hand, the $S_{\mathcal{R}}$ can also be calculated via the \textit{Island formula \uppercase\expandafter{\romannumeral1}} in Weyl transformed CFT$_2$, and we will find configurations with islands which give smaller entanglement entropy than the non-island configurations. This is a result of the finite cutoff introduced by the Weyl transformation. In the following, we will see that the calculation exactly coincides with the holographic results.

\begin{figure}[ht]
	\centering
	\includegraphics[scale=0.4]{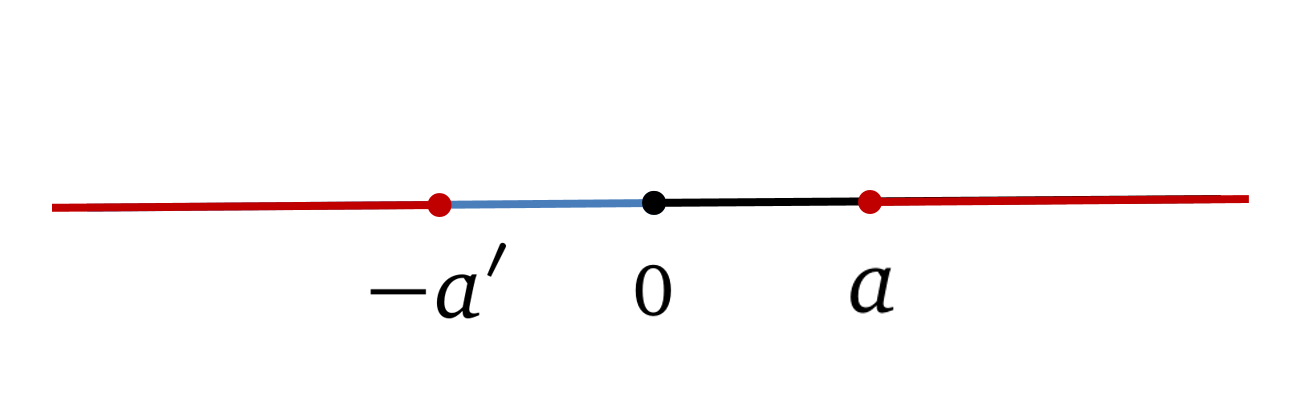}\quad \includegraphics[scale=0.4]{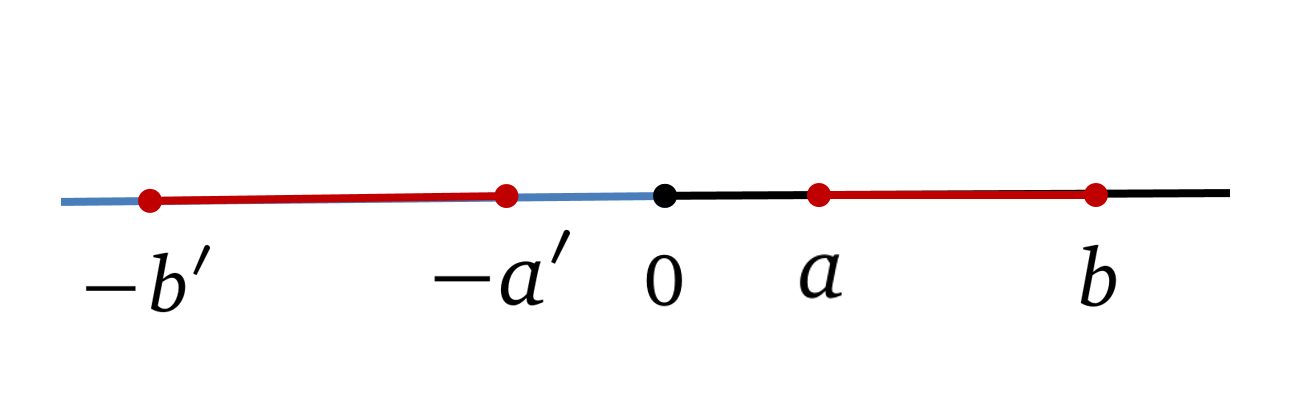}
	\caption{Two configurations in a Weyl-transformed CFT$_2$ where the island formula may be realized: (left) the entanglement entropy of $\mathcal{R}=[a,\infty)$ in the region $x>0$, admits a minimal saddle by including the island $I=(-\infty,-a']$, (right) $\mathcal{R}=[a,b]$ acquires the island $I=[-b',-a']$.}
	\label{fig:islandsinCFT}
\end{figure}

Firstly, let us consider $\mathcal{R}$ to be the semi-infinite region $x>a$ (see the left figure in Fig.\ref{fig:islandsinCFT}) and apply the \textit{Island formula \uppercase\expandafter{\romannumeral1}} and \eqref{entropyweyl} to calculate the entanglement entropy. The entanglement entropy calculated by the island formula is given by
\begin{align}
	S_{\mathcal{R}}=\text{min}\left\{\frac{c}{3}\log\frac{a+a'}{\delta}-\frac{c}{6}\log\left(\frac{2a'}{\delta}\right)+\frac{c}{6}\kappa \right\}\,,\label{island-comp-1}
\end{align}
where we have admitted the region $I=(-\infty,-a']$ as the island corresponding to $\mathcal{R}$. The entropy in \cref{island-comp-1}
has a minimal saddle point
\begin{align}
	S_{\mathcal{R}}=\frac{c}{6}  \log \left(\frac{2 a}{\delta }\right)+\frac{c}{6}\kappa\,,\qquad a'=a\,.
\end{align}
This result is smaller than the entanglement entropy in the non-island phase, hence the entanglement island emerges. Note that, the area term $\frac{Area(X)}{4G}$ does not appear since we assume that the gravitational theory on the $x<0$ region is a induced gravity (see also \cite{Suzuki:2022xwv}). 

Similarly when we consider $\mathcal{R}$ to be an interval $[a,b]$ inside the region $x>0$ and include the corresponding island $I=[-b',-a']$ (see the right figure in Fig.\ref{fig:islandsinCFT}), the \textit{Island formula \uppercase\expandafter{\romannumeral1}} will give\footnote{Here we need to assume that the entanglement entropy for two disjoint intervals in the holographic Weyl transformed CFT exhibit similar phase transitions as the RT formula \cite{Hartman:2013mia,Sully:2020pza}, under certain sparseness
	conditions on the spectrum and OPE coefficients of bulk and boundary operators and large $c$ limit. We leave this for future investigation. }
\begin{align}
	S_{\mathcal{R}}=\text{min}\left\{\frac{c}{3}\log\frac{a+a'}{\delta}+\frac{c}{3}\log\frac{b+b'}{\delta}-\frac{c}{6}\log\left(\frac{2a'}{\delta}\right)-\frac{c}{6}\log\left(\frac{2b'}{\delta}\right)+\frac{c}{3}\kappa\right\}\,,
\end{align}
which has the saddle point
\begin{align}\label{SRisland}
	S_{\mathcal{R}}=\frac{c}{6} \log \left(4\frac{ a b}{\delta^2 }\right)+\frac{c}{3} \kappa\,,\qquad a'=a\,,\quad b'=b\,.
\end{align}
Then we compare the entanglement entropy \eqref{SRisland} in island phase with the one in the non-island phase, which is given by
\begin{align}
	S_{\mathcal{R}}=\frac{c}{3}\log\left(\frac{b-a}{\delta}\right)\,.
\end{align} 
We find that when 
\begin{align}
	0<a<b \left(1-2 \sqrt{e^{2 \kappa}+e^{4 \kappa}}+2 e^{2 \kappa}\right)\,,
\end{align}
the entanglement entropy \eqref{SRisland} calculated by the island formula is smaller, hence the configuration enters the island phase.

If we set $\kappa=\rho_0$, the cutoff brane for the holographic Weyl transformed CFT overlaps with the EoW brane in the AdS/BCFT setup. Obviously, in both of the setups the calculations for the entanglement entropy via the \textit{Island formula \uppercase\expandafter{\romannumeral1}} are exactly the same. 

\section{Discussion}
\label{sec:5-discussions}
In this paper we refined the concept of self-encoding property, and relate it directly to the emergence of entanglement islands and island formulas. Recent developments on replica wormholes in gravitational theories indicated that, the self-encoding property, as well as the island formula $I$, is a result of the emergence of spacetime wormholes, which is an intrinsic property of gravitational theories. On the other hand, we proposed a new mechanism that induces self-encoding properties in non-gravitational systems from a purely quantum information perspective. More explicitly, we can impose certain external constraints on the system hence reduce the Hilbert space of the entire system in a proper way, such that certain coding relations emerge in the reduced Hilbert space. We showed explicitly how the self-encoding property changes the way we compute the reduced density matrix and how the island formula $\uppercase\expandafter{\romannumeral2}$ arises when we compute the entanglement entropy in self-encoded systems. Given the self-encoding property with explicit coding-relations, the island formula $I$ and $II$ are actually the same. The difference between the two formulas is whether the self-encoding property comes from intrinsic spacetime wormholes or external constraints. 

The self-encoding property may shed new light on our understanding of quantum gravity. The \textit{effective theory} of a (quantum) gravity may be understand as a reduced Hilbert space of a quantum field theory coupled to gravity. In configurations where quantum effects of gravity are not significant, this field theory could be understood as a field theory in curved geometric background with sub-leading gravitational fluctuations. The Hilbert space reduction should be consistent with the self-encoding property determined by the wormhole structures. In \cite{Akers:2022qdl}, the authors introduced a non-isometric mapping which projected out lots of state in the Hilbert space, which they call the null states, hence the Hilbert space is tremendously reduced to match the dimension of the Hilbert space of the ``fundamental description''. We believe that their non-isometric mapping plays a similar role as the Hilbert space reduction we have discussed which leads to the self-encoding property. The self-encoding property of gravity make the Hilbert not factorizable, as was pointed out in \cite{Raju:2021lwh}, but according to our discussion, this is consistent with the island formula. 	

We used a special Weyl transformation on a holographic CFT$_2$ to simulate the Hilbert space reduction of 2d gravity, which introduces a finite cutoff scale to the theory. On the field theory side, we assumed that the Weyl transformed part of the CFT is coupled to an AdS$_2$ gravity hence the island formula $I$ applies. On the AdS$_3$ gravity side, we introduces the so-called cutoff spheres and cutoff branes to give a geometric description for the Weyl transformation. We find that the entanglement entropy calculated by the island formula $I$ on the CFT side, coincide with the RT formula on the gravity side. Indeed this simulation exactly captures the main features of the AdS/BCFT configurations. Note that, following this idea, the authors in \cite{Lin:2023ajt} simulated the AdS/BCFT configuration including the first order fluctuation of the EoW brane by adjusting the Weyl transformation accordingly. Furthermore, in the fluctuating brane configuration, the correspondence between the entanglement wedge cross-section and the so-called balanced partial entanglement entropy \cite{Basu:2023wmv} is tested under this simulation. More interestingly, it was shown in \cite{Chandra:2024bkn} that, the Weyl transformation characterized by \eqref{varphi1} is the one that optimizes the path-integral computation for the reduced density matrix of the $x>0$ region in the sense of \cite{Caputa:2018xuf}. The scalar field indeed describes the metric of the $x<0$ region after the Weyl transformation, which indicates that the AdS$_2$ geometry induced by the Weyl transformation is a saddle point of certain theory of the scalar field. If we consider the scalar field to be dynamical, this theory is indeed the gravitational theory coupled to the $x<0$ region, which supports our assumption that the $x<0$ region is coupled to an AdS$_2$ gravity. In other words, the path-integral optimized purification for intervals or half lines in CFT$_2$ are states in island phases. This new perspective implies that the path-integral optimized purifications are self-encoded hence explains the emergence of ``negative mutual information'' in such states \cite{Camargo:2022mme}.  It will also be interesting to consider other Weyl transformations to simulate other generalized version of AdS/BCFT, for example the Wedge holography \cite{Akal:2020wfl}.

\subsection*{Data Availability Statement}
No Data is associated in the manuscript.
\section*{Acknowledgements}
We would like to thank Ziming Ji, Rong-xin Miao, Tadashi Takayanagi, Huajia Wang, Zhenbin Yang and especially Hao Geng and Ling-yan Hung very much for very insightful discussions. We also thank Hao Geng and Ling-yan Hung for valuable comments and suggestions on the manuscript. Qiang Wen and Shangjie Zhou are supported by the ``Zhishan'' scholars program of Southeast University.

\appendix
\section{The cutoff spheres and their common tangent}\label{appendix}
The AdS$_3$ metric in Poincar\'e coordinates $(t,x,z)$ is given by 
\begin{align}
	\dd{s^2}=\frac{-\dd{t^2}+\dd{x^2}+\dd{z^2}}{z^2}.
\end{align}
In the light-cone coordinates
\begin{align}
	U=\frac{x+t}{2},\quad V=\frac{x-t}{2},\quad\rho=\frac{2}{z^2},
\end{align}
the length of the geodesic connecting two spacelike-separated points is\cite{Wen:2018mev}
\begin{align}
	L_{\mathrm{AdS}}&(U_1,V_1,\rho_1,U_2,V_2,\rho_2)\notag\\
	&=\frac{1}{2}\log\left[\frac{\rho_2(\rho_2+X)+\rho_1(\rho_2 Y(2\rho_2+X)+X)+(\rho_1+\rho_2\rho_1 Y)^2}{2\rho_1\rho_2}\right]\,,\label{Geod-length-Poincare}
\end{align}
where 
\begin{align}
	\begin{aligned}
		Y=&2(U_1-U_2)(V_1-V_2)\,,\\
		X=&\sqrt[]{\rho_1^2+2\rho_2\rho_1(\rho_1 Y-1)+(\rho_2+\rho_1\rho_2 Y)^2}.
	\end{aligned}
\end{align}

We look for the set of points $(0,x,z)$ whose geodesic distance from a fixed point $(0,x_0,\delta)$ is a constant $\abs{\phi(x_0)}=\log(\frac{2\abs{x_0}}{\delta})-\kappa$. The equation that $x$ and $z$ should satisfy can be obtained straightforwardly by applying \cref{Geod-length-Poincare}:
\begin{align}
	\frac{1}{2}\log\left[\frac{\rho_2(\rho_2+X)+\rho_1(\rho_2 Y(2\rho_2+X)+X)+(\rho_1+\rho_2\rho_1 Y)^2}{2\rho_1\rho_2}\right]=\log(\frac{2\abs{x_0}}{\delta})-\kappa\,,
\end{align}
which, after taking the limit $\delta\to 0$, can be simplified to be,
\begin{align}
	\left(x-{x_0}\right){}^2+\left(z-\abs{x_0} e^{-\kappa }\right)^2=\abs{x_0}^2 e^{-2\kappa } \,,
\end{align}
which is a circle at $(x_0,\abs{x_0} e^{-\kappa })$ with radius $r=\abs{x_0} e^{-\kappa } $.

\begin{figure}[ht]
	\centering
	\includegraphics[scale=0.55]{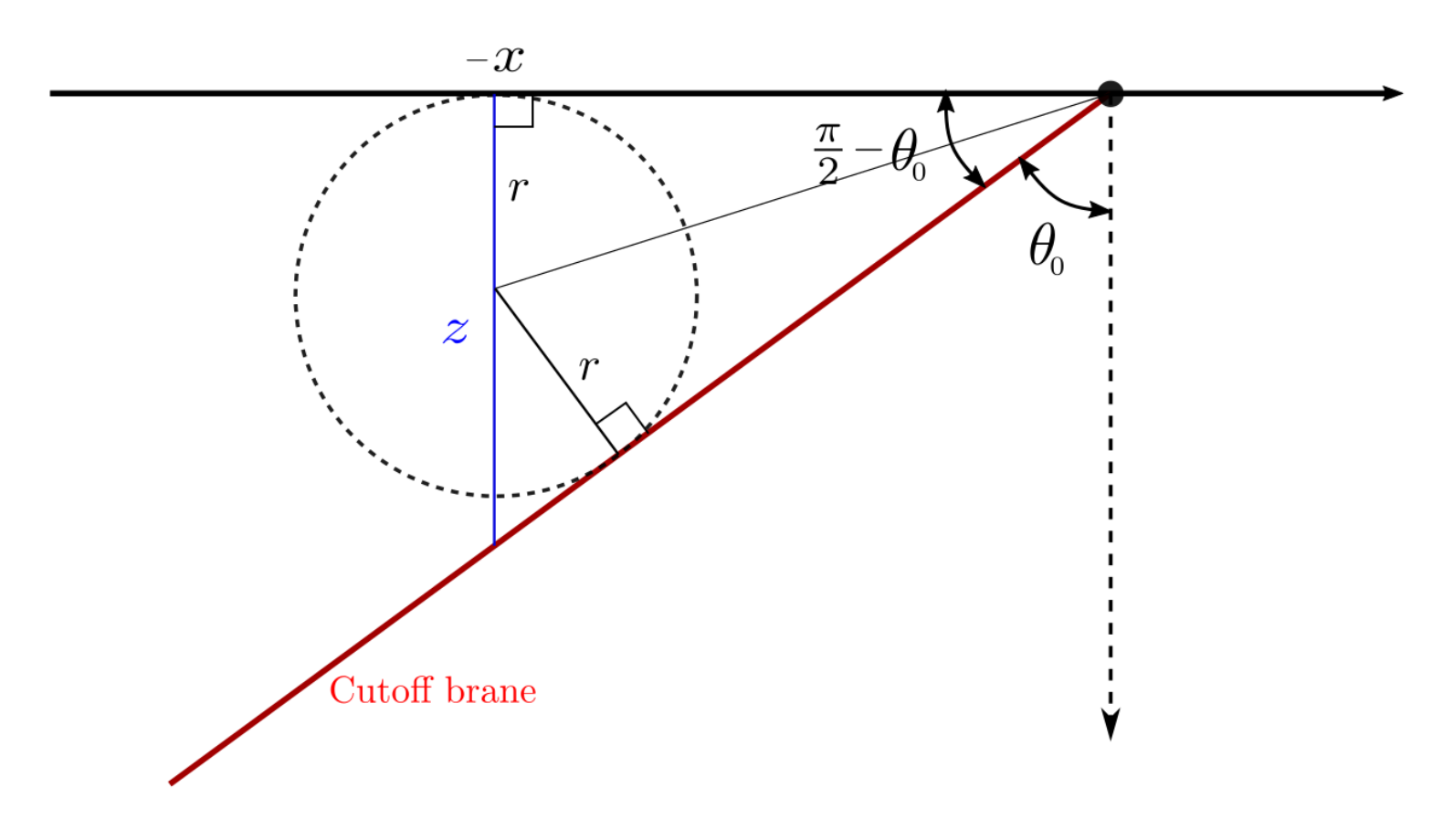}
	\caption{A generic cutoff sphere at $-x$ with radius $r=xe^{-\kappa}$ is centered in the bulk at the point $(x,xe^{-\kappa})$. The tangent from $x=0$ shown by red line acts as a cutoff brane which is equivalent to the EoW brane in the AdS/BCFT scenario.}
	\label{fig:cut-off1}
\end{figure}
In \cref{fig:cut-off1}, a generic cut-off sphere at the point $-x$ (x>0) with radius $r=x e^{-\kappa }$ is depicted. The tangent to the cut-off sphere from $x=0$ is shown by the red line. We may obtain the angle $\theta_0$ of the tangent line with the vertical as follows
\begin{align}
	\tan\left(\frac{\pi}{4}-\frac{\theta_0}{2}\right)=\frac{r}{|x|}=e^{-\kappa}\,.
\end{align}
Hence, the hyperbolic angle $\rho$ for the tangent line is obtained as
\begin{align}
	\rho=\text{arccosh}\left(\frac{1}{\cos\theta_0}\right)=\kappa\,,
\end{align}
which confirms our claim that the cutoff brane obtained from the common tangent line of all the cut-off spheres is equivalent to the end-of-the-world brane in the AdS/BCFT setup.

	\bibliography{reference}

\begin{thebibliography}{100}
\providecommand{\url}[1]{\texttt{#1}}
\providecommand{\urlprefix}{URL }
\expandafter\ifx\csname urlstyle\endcsname\relax
  \providecommand{\doi}[1]{doi:\discretionary{}{}{}#1}\else
  \providecommand{\doi}{doi:\discretionary{}{}{}\begingroup
  \urlstyle{rm}\Url}\fi
\providecommand{\eprint}[2][]{\url{#2}}

\bibitem{Hawking:1976ra}
S.~W. Hawking,
\newblock \emph{{Breakdown of Predictability in Gravitational Collapse}},
\newblock Phys. Rev. D \textbf{14}, 2460 (1976),
\newblock \doi{10.1103/PhysRevD.14.2460}.

\bibitem{Page:1993wv}
D.~N. Page,
\newblock \emph{{Information in black hole radiation}},
\newblock Phys. Rev. Lett. \textbf{71}, 3743 (1993),
\newblock \doi{10.1103/PhysRevLett.71.3743},
\newblock \eprint{hep-th/9306083}.

\bibitem{Page:2013dx}
D.~N. Page,
\newblock \emph{{Time Dependence of Hawking Radiation Entropy}},
\newblock JCAP \textbf{09}, 028 (2013),
\newblock \doi{10.1088/1475-7516/2013/09/028},
\newblock \eprint{1301.4995}.

\bibitem{Mathur:2009hf}
S.~D. Mathur,
\newblock \emph{{The Information paradox: A Pedagogical introduction}},
\newblock Class. Quant. Grav. \textbf{26}, 224001 (2009),
\newblock \doi{10.1088/0264-9381/26/22/224001},
\newblock \eprint{0909.1038}.

\bibitem{Almheiri:2012rt}
A.~Almheiri, D.~Marolf, J.~Polchinski and J.~Sully,
\newblock \emph{{Black Holes: Complementarity or Firewalls?}},
\newblock JHEP \textbf{02}, 062 (2013),
\newblock \doi{10.1007/JHEP02(2013)062},
\newblock \eprint{1207.3123}.

\bibitem{Maldacena:1997re}
J.~M. Maldacena,
\newblock \emph{{The Large N limit of superconformal field theories and
  supergravity}},
\newblock Adv. Theor. Math. Phys. \textbf{2}, 231 (1998),
\newblock \doi{10.1023/A:1026654312961},
\newblock \eprint{hep-th/9711200}.

\bibitem{Gubser:1998bc}
S.~S. Gubser, I.~R. Klebanov and A.~M. Polyakov,
\newblock \emph{{Gauge theory correlators from noncritical string theory}},
\newblock Phys. Lett. B \textbf{428}, 105 (1998),
\newblock \doi{10.1016/S0370-2693(98)00377-3},
\newblock \eprint{hep-th/9802109}.

\bibitem{Witten:1998qj}
E.~Witten,
\newblock \emph{{Anti-de Sitter space and holography}},
\newblock Adv. Theor. Math. Phys. \textbf{2}, 253 (1998),
\newblock \doi{10.4310/ATMP.1998.v2.n2.a2},
\newblock \eprint{hep-th/9802150}.

\bibitem{Ryu:2006bv}
S.~Ryu and T.~Takayanagi,
\newblock \emph{{Holographic derivation of entanglement entropy from AdS/CFT}},
\newblock Phys. Rev. Lett. \textbf{96}, 181602 (2006),
\newblock \doi{10.1103/PhysRevLett.96.181602},
\newblock \eprint{hep-th/0603001}.

\bibitem{Ryu:2006ef}
S.~Ryu and T.~Takayanagi,
\newblock \emph{{Aspects of Holographic Entanglement Entropy}},
\newblock JHEP \textbf{08}, 045 (2006),
\newblock \doi{10.1088/1126-6708/2006/08/045},
\newblock \eprint{hep-th/0605073}.

\bibitem{Lewkowycz:2013nqa}
A.~Lewkowycz and J.~Maldacena,
\newblock \emph{{Generalized gravitational entropy}},
\newblock JHEP \textbf{08}, 090 (2013),
\newblock \doi{10.1007/JHEP08(2013)090},
\newblock \eprint{1304.4926}.

\bibitem{Faulkner:2013ana}
T.~Faulkner, A.~Lewkowycz and J.~Maldacena,
\newblock \emph{{Quantum corrections to holographic entanglement entropy}},
\newblock JHEP \textbf{11}, 074 (2013),
\newblock \doi{10.1007/JHEP11(2013)074},
\newblock \eprint{1307.2892}.

\bibitem{Engelhardt:2014gca}
N.~Engelhardt and A.~C. Wall,
\newblock \emph{{Quantum Extremal Surfaces: Holographic Entanglement Entropy
  beyond the Classical Regime}},
\newblock JHEP \textbf{01}, 073 (2015),
\newblock \doi{10.1007/JHEP01(2015)073},
\newblock \eprint{1408.3203}.

\bibitem{Wall:2012uf}
A.~C. Wall,
\newblock \emph{{Maximin Surfaces, and the Strong Subadditivity of the
  Covariant Holographic Entanglement Entropy}},
\newblock Class. Quant. Grav. \textbf{31}(22), 225007 (2014),
\newblock \doi{10.1088/0264-9381/31/22/225007},
\newblock \eprint{1211.3494}.

\bibitem{Dong:2016hjy}
X.~Dong, A.~Lewkowycz and M.~Rangamani,
\newblock \emph{{Deriving covariant holographic entanglement}},
\newblock JHEP \textbf{11}, 028 (2016),
\newblock \doi{10.1007/JHEP11(2016)028},
\newblock \eprint{1607.07506}.

\bibitem{Akers:2019lzs}
C.~Akers, N.~Engelhardt, G.~Penington and M.~Usatyuk,
\newblock \emph{{Quantum Maximin Surfaces}},
\newblock JHEP \textbf{08}, 140 (2020),
\newblock \doi{10.1007/JHEP08(2020)140},
\newblock \eprint{1912.02799}.

\bibitem{Dong:2017xht}
X.~Dong and A.~Lewkowycz,
\newblock \emph{{Entropy, Extremality, Euclidean Variations, and the Equations
  of Motion}},
\newblock JHEP \textbf{01}, 081 (2018),
\newblock \doi{10.1007/JHEP01(2018)081},
\newblock \eprint{1705.08453}.

\bibitem{Hubeny:2007xt}
V.~E. Hubeny, M.~Rangamani and T.~Takayanagi,
\newblock \emph{{A Covariant holographic entanglement entropy proposal}},
\newblock JHEP \textbf{07}, 062 (2007),
\newblock \doi{10.1088/1126-6708/2007/07/062},
\newblock \eprint{0705.0016}.

\bibitem{Penington:2019npb}
G.~Penington,
\newblock \emph{{Entanglement Wedge Reconstruction and the Information
  Paradox}},
\newblock JHEP \textbf{09}, 002 (2020),
\newblock \doi{10.1007/JHEP09(2020)002},
\newblock \eprint{1905.08255}.

\bibitem{Almheiri:2019psf}
A.~Almheiri, N.~Engelhardt, D.~Marolf and H.~Maxfield,
\newblock \emph{{The entropy of bulk quantum fields and the entanglement wedge
  of an evaporating black hole}},
\newblock JHEP \textbf{12}, 063 (2019),
\newblock \doi{10.1007/JHEP12(2019)063},
\newblock \eprint{1905.08762}.

\bibitem{Rozali:2019day}
M.~Rozali, J.~Sully, M.~Van~Raamsdonk, C.~Waddell and D.~Wakeham,
\newblock \emph{{Information radiation in BCFT models of black holes}},
\newblock JHEP \textbf{05}, 004 (2020),
\newblock \doi{10.1007/JHEP05(2020)004},
\newblock \eprint{1910.12836}.

\bibitem{Chen:2019uhq}
H.~Z. Chen, Z.~Fisher, J.~Hernandez, R.~C. Myers and S.-M. Ruan,
\newblock \emph{{Information Flow in Black Hole Evaporation}},
\newblock JHEP \textbf{03}, 152 (2020),
\newblock \doi{10.1007/JHEP03(2020)152},
\newblock \eprint{1911.03402}.

\bibitem{Almheiri:2019hni}
A.~Almheiri, R.~Mahajan, J.~Maldacena and Y.~Zhao,
\newblock \emph{{The Page curve of Hawking radiation from semiclassical
  geometry}},
\newblock JHEP \textbf{03}, 149 (2020),
\newblock \doi{10.1007/JHEP03(2020)149},
\newblock \eprint{1908.10996}.

\bibitem{Almheiri:2019psy}
A.~Almheiri, R.~Mahajan and J.~E. Santos,
\newblock \emph{{Entanglement islands in higher dimensions}},
\newblock SciPost Phys. \textbf{9}(1), 001 (2020),
\newblock \doi{10.21468/SciPostPhys.9.1.001},
\newblock \eprint{1911.09666}.

\bibitem{Almheiri:2019yqk}
A.~Almheiri, R.~Mahajan and J.~Maldacena,
\newblock \emph{{Islands outside the horizon}}  (2019),
\newblock \eprint{1910.11077}.

\bibitem{Almheiri:2019qdq}
A.~Almheiri, T.~Hartman, J.~Maldacena, E.~Shaghoulian and A.~Tajdini,
\newblock \emph{{Replica Wormholes and the Entropy of Hawking Radiation}},
\newblock JHEP \textbf{05}, 013 (2020),
\newblock \doi{10.1007/JHEP05(2020)013},
\newblock \eprint{1911.12333}.

\bibitem{Penington:2019kki}
G.~Penington, S.~H. Shenker, D.~Stanford and Z.~Yang,
\newblock \emph{{Replica wormholes and the black hole interior}},
\newblock JHEP \textbf{03}, 205 (2022),
\newblock \doi{10.1007/JHEP03(2022)205},
\newblock \eprint{1911.11977}.

\bibitem{Takayanagi:2011zk}
T.~Takayanagi,
\newblock \emph{{Holographic Dual of BCFT}},
\newblock Phys. Rev. Lett. \textbf{107}, 101602 (2011),
\newblock \doi{10.1103/PhysRevLett.107.101602},
\newblock \eprint{1105.5165}.

\bibitem{Chen:2019iro}
Y.~Chen,
\newblock \emph{{Pulling Out the Island with Modular Flow}},
\newblock JHEP \textbf{03}, 033 (2020),
\newblock \doi{10.1007/JHEP03(2020)033},
\newblock \eprint{1912.02210}.

\bibitem{Chen:2020wiq}
Y.~Chen, X.-L. Qi and P.~Zhang,
\newblock \emph{{Replica wormhole and information retrieval in the SYK model
  coupled to Majorana chains}},
\newblock JHEP \textbf{06}, 121 (2020),
\newblock \doi{10.1007/JHEP06(2020)121},
\newblock \eprint{2003.13147}.

\bibitem{Chen:2020hmv}
H.~Z. Chen, R.~C. Myers, D.~Neuenfeld, I.~A. Reyes and J.~Sandor,
\newblock \emph{{Quantum Extremal Islands Made Easy, Part II: Black Holes on
  the Brane}},
\newblock JHEP \textbf{12}, 025 (2020),
\newblock \doi{10.1007/JHEP12(2020)025},
\newblock \eprint{2010.00018}.

\bibitem{Hernandez:2020nem}
J.~Hernandez, R.~C. Myers and S.-M. Ruan,
\newblock \emph{{Quantum extremal islands made easy. Part III. Complexity on
  the brane}},
\newblock JHEP \textbf{02}, 173 (2021),
\newblock \doi{10.1007/JHEP02(2021)173},
\newblock \eprint{2010.16398}.

\bibitem{Grimaldi:2022suv}
G.~Grimaldi, J.~Hernandez and R.~C. Myers,
\newblock \emph{{Quantum extremal islands made easy. Part IV. Massive black
  holes on the brane}},
\newblock JHEP \textbf{03}, 136 (2022),
\newblock \doi{10.1007/JHEP03(2022)136},
\newblock \eprint{2202.00679}.

\bibitem{Akal:2020twv}
I.~Akal, Y.~Kusuki, N.~Shiba, T.~Takayanagi and Z.~Wei,
\newblock \emph{{Entanglement Entropy in a Holographic Moving Mirror and the
  Page Curve}},
\newblock Phys. Rev. Lett. \textbf{126}(6), 061604 (2021),
\newblock \doi{10.1103/PhysRevLett.126.061604},
\newblock \eprint{2011.12005}.

\bibitem{Deng:2020ent}
F.~Deng, J.~Chu and Y.~Zhou,
\newblock \emph{{Defect extremal surface as the holographic counterpart of
  Island formula}},
\newblock JHEP \textbf{03}, 008 (2021),
\newblock \doi{10.1007/JHEP03(2021)008},
\newblock \eprint{2012.07612}.

\bibitem{Anous:2022wqh}
T.~Anous, M.~Meineri, P.~Pelliconi and J.~Sonner,
\newblock \emph{{Sailing past the End of the World and discovering the
  Island}},
\newblock SciPost Phys. \textbf{13}(3), 075 (2022),
\newblock \doi{10.21468/SciPostPhys.13.3.075},
\newblock \eprint{2202.11718}.

\bibitem{Geng:2020qvw}
H.~Geng and A.~Karch,
\newblock \emph{{Massive islands}},
\newblock JHEP \textbf{09}, 121 (2020),
\newblock \doi{10.1007/JHEP09(2020)121},
\newblock \eprint{2006.02438}.

\bibitem{Karlsson:2020uga}
A.~Karlsson,
\newblock \emph{{Replica wormhole and island incompatibility with monogamy of
  entanglement}}  (2020),
\newblock \eprint{2007.10523}.

\bibitem{Raju:2020smc}
S.~Raju,
\newblock \emph{{Lessons from the information paradox}},
\newblock Phys. Rept. \textbf{943}, 2187 (2022),
\newblock \doi{10.1016/j.physrep.2021.10.001},
\newblock \eprint{2012.05770}.

\bibitem{Raju:2021lwh}
S.~Raju,
\newblock \emph{{Failure of the split property in gravity and the information
  paradox}},
\newblock Class. Quant. Grav. \textbf{39}(6), 064002 (2022),
\newblock \doi{10.1088/1361-6382/ac482b},
\newblock \eprint{2110.05470}.

\bibitem{Laddha:2020kvp}
A.~Laddha, S.~G. Prabhu, S.~Raju and P.~Shrivastava,
\newblock \emph{{The Holographic Nature of Null Infinity}},
\newblock SciPost Phys. \textbf{10}(2), 041 (2021),
\newblock \doi{10.21468/SciPostPhys.10.2.041},
\newblock \eprint{2002.02448}.

\bibitem{Geng:2020fxl}
H.~Geng, A.~Karch, C.~Perez-Pardavila, S.~Raju, L.~Randall, M.~Riojas and
  S.~Shashi,
\newblock \emph{{Information Transfer with a Gravitating Bath}},
\newblock SciPost Phys. \textbf{10}(5), 103 (2021),
\newblock \doi{10.21468/SciPostPhys.10.5.103},
\newblock \eprint{2012.04671}.

\bibitem{Geng:2021hlu}
H.~Geng, A.~Karch, C.~Perez-Pardavila, S.~Raju, L.~Randall, M.~Riojas and
  S.~Shashi,
\newblock \emph{{Inconsistency of islands in theories with long-range
  gravity}},
\newblock JHEP \textbf{01}, 182 (2022),
\newblock \doi{10.1007/JHEP01(2022)182},
\newblock \eprint{2107.03390}.

\bibitem{Alishahiha:2020qza}
M.~Alishahiha, A.~Faraji~Astaneh and A.~Naseh,
\newblock \emph{{Island in the presence of higher derivative terms}},
\newblock JHEP \textbf{02}, 035 (2021),
\newblock \doi{10.1007/JHEP02(2021)035},
\newblock \eprint{2005.08715}.

\bibitem{HosseiniMansoori:2022hok}
S.~A. Hosseini~Mansoori, O.~Luongo, S.~Mancini, M.~Mirjalali, M.~Rafiee and
  A.~Tavanfar,
\newblock \emph{{Planar black holes in holographic axion gravity: Islands, Page
  times, and scrambling times}},
\newblock Phys. Rev. D \textbf{106}(12), 126018 (2022),
\newblock \doi{10.1103/PhysRevD.106.126018},
\newblock \eprint{2209.00253}.

\bibitem{Karch:2022rvr}
A.~Karch, H.~Sun and C.~F. Uhlemann,
\newblock \emph{{Double holography in string theory}},
\newblock JHEP \textbf{10}, 012 (2022),
\newblock \doi{10.1007/JHEP10(2022)012},
\newblock \eprint{2206.11292}.

\bibitem{Uhlemann:2021nhu}
C.~F. Uhlemann,
\newblock \emph{{Islands and Page curves in 4d from Type IIB}},
\newblock JHEP \textbf{08}, 104 (2021),
\newblock \doi{10.1007/JHEP08(2021)104},
\newblock \eprint{2105.00008}.

\bibitem{Krishnan:2020oun}
C.~Krishnan, V.~Patil and J.~Pereira,
\newblock \emph{{Page Curve and the Information Paradox in Flat Space}}
  (2020),
\newblock \eprint{2005.02993}.

\bibitem{Krishnan:2020fer}
C.~Krishnan,
\newblock \emph{{Critical Islands}},
\newblock JHEP \textbf{01}, 179 (2021),
\newblock \doi{10.1007/JHEP01(2021)179},
\newblock \eprint{2007.06551}.

\bibitem{Ghosh:2021axl}
K.~Ghosh and C.~Krishnan,
\newblock \emph{{Dirichlet baths and the not-so-fine-grained Page curve}},
\newblock JHEP \textbf{08}, 119 (2021),
\newblock \doi{10.1007/JHEP08(2021)119},
\newblock \eprint{2103.17253}.

\bibitem{Almheiri:2020cfm}
A.~Almheiri, T.~Hartman, J.~Maldacena, E.~Shaghoulian and A.~Tajdini,
\newblock \emph{{The entropy of Hawking radiation}},
\newblock Rev. Mod. Phys. \textbf{93}(3), 035002 (2021),
\newblock \doi{10.1103/RevModPhys.93.035002},
\newblock \eprint{2006.06872}.

\bibitem{Bousso:2022ntt}
R.~Bousso, X.~Dong, N.~Engelhardt, T.~Faulkner, T.~Hartman, S.~H. Shenker and
  D.~Stanford,
\newblock \emph{{Snowmass White Paper: Quantum Aspects of Black Holes and the
  Emergence of Spacetime}}  (2022),
\newblock \eprint{2201.03096}.

\bibitem{Miao:2022mdx}
R.-X. Miao,
\newblock \emph{{Massless Entanglement Island in Wedge Holography}}  (2022),
\newblock \eprint{2212.07645}.

\bibitem{Miao:2023unv}
R.-X. Miao,
\newblock \emph{{Entanglement island and Page curve in wedge holography}},
\newblock JHEP \textbf{03}, 214 (2023),
\newblock \doi{10.1007/JHEP03(2023)214},
\newblock \eprint{2301.06285}.

\bibitem{Renner:2021qbe}
R.~Renner and J.~Wang,
\newblock \emph{{The black hole information puzzle and the quantum de Finetti
  theorem}}  (2021),
\newblock \eprint{2110.14653}.

\bibitem{Wang:2021afl}
X.~Wang, K.~Zhang and J.~Wang,
\newblock \emph{{What can we learn about islands and state paradox from quantum
  information theory?}}  (2021),
\newblock \eprint{2107.09228}.

\bibitem{Akers:2022qdl}
C.~Akers, N.~Engelhardt, D.~Harlow, G.~Penington and S.~Vardhan,
\newblock \emph{{The black hole interior from non-isometric codes and
  complexity}}  (2022),
\newblock \eprint{2207.06536}.

\bibitem{Almheiri:2021jwq}
A.~Almheiri and H.~W. Lin,
\newblock \emph{{The entanglement wedge of unknown couplings}},
\newblock JHEP \textbf{08}, 062 (2022),
\newblock \doi{10.1007/JHEP08(2022)062},
\newblock \eprint{2111.06298}.

\bibitem{1993QuantumEA}
M.~Ohya and D.~Petz,
\newblock \emph{Quantum entropy and its use} (1993).

\bibitem{Calabrese:2004eu}
P.~Calabrese and J.~L. Cardy,
\newblock \emph{{Entanglement entropy and quantum field theory}},
\newblock J. Stat. Mech. \textbf{0406}, P06002 (2004),
\newblock \doi{10.1088/1742-5468/2004/06/P06002},
\newblock \eprint{hep-th/0405152}.

\bibitem{Calabrese:2009qy}
P.~Calabrese and J.~Cardy,
\newblock \emph{{Entanglement entropy and conformal field theory}},
\newblock J. Phys. A \textbf{42}, 504005 (2009),
\newblock \doi{10.1088/1751-8113/42/50/504005},
\newblock \eprint{0905.4013}.

\bibitem{Casini:2013rba}
H.~Casini, M.~Huerta and J.~A. Rosabal,
\newblock \emph{{Remarks on entanglement entropy for gauge fields}},
\newblock Phys. Rev. D \textbf{89}(8), 085012 (2014),
\newblock \doi{10.1103/PhysRevD.89.085012},
\newblock \eprint{1312.1183}.

\bibitem{Ghosh:2015iwa}
S.~Ghosh, R.~M. Soni and S.~P. Trivedi,
\newblock \emph{{On The Entanglement Entropy For Gauge Theories}},
\newblock JHEP \textbf{09}, 069 (2015),
\newblock \doi{10.1007/JHEP09(2015)069},
\newblock \eprint{1501.02593}.

\bibitem{Hung:2015fla}
L.-Y. Hung and Y.~Wan,
\newblock \emph{{Revisiting Entanglement Entropy of Lattice Gauge Theories}},
\newblock JHEP \textbf{04}, 122 (2015),
\newblock \doi{10.1007/JHEP04(2015)122},
\newblock \eprint{1501.04389}.

\bibitem{Dvali:2008jb}
G.~Dvali and S.~N. Solodukhin,
\newblock \emph{{Black Hole Entropy and Gravity Cutoff}}  (2008),
\newblock \eprint{0806.3976}.

\bibitem{Solodukhin:2011gn}
S.~N. Solodukhin,
\newblock \emph{{Entanglement entropy of black holes}},
\newblock Living Rev. Rel. \textbf{14}, 8 (2011),
\newblock \doi{10.12942/lrr-2011-8},
\newblock \eprint{1104.3712}.

\bibitem{Hawking:2000da}
S.~Hawking, J.~M. Maldacena and A.~Strominger,
\newblock \emph{{de Sitter entropy, quantum entanglement and AdS / CFT}},
\newblock JHEP \textbf{05}, 001 (2001),
\newblock \doi{10.1088/1126-6708/2001/05/001},
\newblock \eprint{hep-th/0002145}.

\bibitem{Dvali:2007hz}
G.~Dvali,
\newblock \emph{{Black Holes and Large N Species Solution to the Hierarchy
  Problem}},
\newblock Fortsch. Phys. \textbf{58}, 528 (2010),
\newblock \doi{10.1002/prop.201000009},
\newblock \eprint{0706.2050}.

\bibitem{Bousso:2012as}
R.~Bousso,
\newblock \emph{{Complementarity Is Not Enough}},
\newblock Phys. Rev. D \textbf{87}(12), 124023 (2013),
\newblock \doi{10.1103/PhysRevD.87.124023},
\newblock \eprint{1207.5192}.

\bibitem{Verlinde:2012cy}
E.~Verlinde and H.~Verlinde,
\newblock \emph{{Black Hole Entanglement and Quantum Error Correction}},
\newblock JHEP \textbf{10}, 107 (2013),
\newblock \doi{10.1007/JHEP10(2013)107},
\newblock \eprint{1211.6913}.

\bibitem{Papadodimas:2012aq}
K.~Papadodimas and S.~Raju,
\newblock \emph{{An Infalling Observer in AdS/CFT}},
\newblock JHEP \textbf{10}, 212 (2013),
\newblock \doi{10.1007/JHEP10(2013)212},
\newblock \eprint{1211.6767}.

\bibitem{Susskind:2013tg}
L.~Susskind,
\newblock \emph{{Black Hole Complementarity and the Harlow-Hayden Conjecture}}
  (2013),
\newblock \eprint{1301.4505}.

\bibitem{Maldacena:2013xja}
J.~Maldacena and L.~Susskind,
\newblock \emph{{Cool horizons for entangled black holes}},
\newblock Fortsch. Phys. \textbf{61}, 781 (2013),
\newblock \doi{10.1002/prop.201300020},
\newblock \eprint{1306.0533}.

\bibitem{Bousso:2012mh}
R.~Bousso, B.~Freivogel, S.~Leichenauer, V.~Rosenhaus and C.~Zukowski,
\newblock \emph{{Null Geodesics, Local CFT Operators and AdS/CFT for
  Subregions}},
\newblock Phys. Rev. D \textbf{88}, 064057 (2013),
\newblock \doi{10.1103/PhysRevD.88.064057},
\newblock \eprint{1209.4641}.

\bibitem{Czech:2012bh}
B.~Czech, J.~L. Karczmarek, F.~Nogueira and M.~Van~Raamsdonk,
\newblock \emph{{The Gravity Dual of a Density Matrix}},
\newblock Class. Quant. Grav. \textbf{29}, 155009 (2012),
\newblock \doi{10.1088/0264-9381/29/15/155009},
\newblock \eprint{1204.1330}.

\bibitem{Headrick:2014cta}
M.~Headrick, V.~E. Hubeny, A.~Lawrence and M.~Rangamani,
\newblock \emph{{Causality \& holographic entanglement entropy}},
\newblock JHEP \textbf{12}, 162 (2014),
\newblock \doi{10.1007/JHEP12(2014)162},
\newblock \eprint{1408.6300}.

\bibitem{Almheiri:2014lwa}
A.~Almheiri, X.~Dong and D.~Harlow,
\newblock \emph{{Bulk Locality and Quantum Error Correction in AdS/CFT}},
\newblock JHEP \textbf{04}, 163 (2015),
\newblock \doi{10.1007/JHEP04(2015)163},
\newblock \eprint{1411.7041}.

\bibitem{Jafferis:2015del}
D.~L. Jafferis, A.~Lewkowycz, J.~Maldacena and S.~J. Suh,
\newblock \emph{{Relative entropy equals bulk relative entropy}},
\newblock JHEP \textbf{06}, 004 (2016),
\newblock \doi{10.1007/JHEP06(2016)004},
\newblock \eprint{1512.06431}.

\bibitem{Dong:2016eik}
X.~Dong, D.~Harlow and A.~C. Wall,
\newblock \emph{{Reconstruction of Bulk Operators within the Entanglement Wedge
  in Gauge-Gravity Duality}},
\newblock Phys. Rev. Lett. \textbf{117}(2), 021601 (2016),
\newblock \doi{10.1103/PhysRevLett.117.021601},
\newblock \eprint{1601.05416}.

\bibitem{Harlow:2016vwg}
D.~Harlow,
\newblock \emph{{The Ryu\textendash{}Takayanagi Formula from Quantum Error
  Correction}},
\newblock Commun. Math. Phys. \textbf{354}(3), 865 (2017),
\newblock \doi{10.1007/s00220-017-2904-z},
\newblock \eprint{1607.03901}.

\bibitem{Faulkner:2017vdd}
T.~Faulkner and A.~Lewkowycz,
\newblock \emph{{Bulk locality from modular flow}},
\newblock JHEP \textbf{07}, 151 (2017),
\newblock \doi{10.1007/JHEP07(2017)151},
\newblock \eprint{1704.05464}.

\bibitem{Petz:1986tvy}
D.~Petz,
\newblock \emph{{Sufficient subalgebras and the relative entropy of states of a
  von Neumann algebra}},
\newblock Commun. Math. Phys. \textbf{105}(1), 123 (1986),
\newblock \doi{10.1007/BF01212345}.

\bibitem{Petz:1988usv}
D.~Petz,
\newblock \emph{{SUFFICIENCY OF CHANNELS OVER VON NEUMANN ALGEBRAS}},
\newblock Quart. J. Math. Oxford Ser. \textbf{39}(1), 97 (1988),
\newblock \doi{10.1093/qmath/39.1.97}.

\bibitem{Cotler:2017erl}
J.~Cotler, P.~Hayden, G.~Penington, G.~Salton, B.~Swingle and M.~Walter,
\newblock \emph{{Entanglement Wedge Reconstruction via Universal Recovery
  Channels}},
\newblock Phys. Rev. X \textbf{9}(3), 031011 (2019),
\newblock \doi{10.1103/PhysRevX.9.031011},
\newblock \eprint{1704.05839}.

\bibitem{Chen:2019gbt}
C.-F. Chen, G.~Penington and G.~Salton,
\newblock \emph{{Entanglement Wedge Reconstruction using the Petz Map}},
\newblock JHEP \textbf{01}, 168 (2020),
\newblock \doi{10.1007/JHEP01(2020)168},
\newblock \eprint{1902.02844}.

\bibitem{Chowdhury:2020hse}
C.~Chowdhury, O.~Papadoulaki and S.~Raju,
\newblock \emph{{A physical protocol for observers near the boundary to obtain
  bulk information in quantum gravity}},
\newblock SciPost Phys. \textbf{10}(5), 106 (2021),
\newblock \doi{10.21468/SciPostPhys.10.5.106},
\newblock \eprint{2008.01740}.

\bibitem{Chowdhury:2021nxw}
C.~Chowdhury, V.~Godet, O.~Papadoulaki and S.~Raju,
\newblock \emph{{Holography from the Wheeler-DeWitt equation}},
\newblock JHEP \textbf{03}, 019 (2022),
\newblock \doi{10.1007/JHEP03(2022)019},
\newblock \eprint{2107.14802}.

\bibitem{Suzuki:2022xwv}
K.~Suzuki and T.~Takayanagi,
\newblock \emph{{BCFT and Islands in two dimensions}},
\newblock JHEP \textbf{06}, 095 (2022),
\newblock \doi{10.1007/JHEP06(2022)095},
\newblock \eprint{2202.08462}.

\bibitem{Caputa:2017urj}
P.~Caputa, N.~Kundu, M.~Miyaji, T.~Takayanagi and K.~Watanabe,
\newblock \emph{{Anti-de Sitter Space from Optimization of Path Integrals in
  Conformal Field Theories}},
\newblock Phys. Rev. Lett. \textbf{119}(7), 071602 (2017),
\newblock \doi{10.1103/PhysRevLett.119.071602},
\newblock \eprint{1703.00456}.

\bibitem{Caputa:2018xuf}
P.~Caputa, M.~Miyaji, T.~Takayanagi and K.~Umemoto,
\newblock \emph{{Holographic Entanglement of Purification from Conformal Field
  Theories}},
\newblock Phys. Rev. Lett. \textbf{122}(11), 111601 (2019),
\newblock \doi{10.1103/PhysRevLett.122.111601},
\newblock \eprint{1812.05268}.

\bibitem{Camargo:2022mme}
H.~A. Camargo, P.~Nandy, Q.~Wen and H.~Zhong,
\newblock \emph{{Balanced partial entanglement and mixed state correlations}},
\newblock SciPost Phys. \textbf{12}(4), 137 (2022),
\newblock \doi{10.21468/SciPostPhys.12.4.137},
\newblock \eprint{2201.13362}.

\bibitem{Fujita:2011fp}
M.~Fujita, T.~Takayanagi and E.~Tonni,
\newblock \emph{{Aspects of AdS/BCFT}},
\newblock JHEP \textbf{11}, 043 (2011),
\newblock \doi{10.1007/JHEP11(2011)043},
\newblock \eprint{1108.5152}.

\bibitem{Sully:2020pza}
J.~Sully, M.~V. Raamsdonk and D.~Wakeham,
\newblock \emph{{BCFT entanglement entropy at large central charge and the
  black hole interior}},
\newblock JHEP \textbf{03}, 167 (2021),
\newblock \doi{10.1007/JHEP03(2021)167},
\newblock \eprint{2004.13088}.

\bibitem{Swingle:2010jz}
B.~Swingle,
\newblock \emph{{Mutual information and the structure of entanglement in
  quantum field theory}}  (2010),
\newblock \eprint{1010.4038}.

\bibitem{Dvali:2000hr}
G.~R. Dvali, G.~Gabadadze and M.~Porrati,
\newblock \emph{{4-D gravity on a brane in 5-D Minkowski space}},
\newblock Phys. Lett. B \textbf{485}, 208 (2000),
\newblock \doi{10.1016/S0370-2693(00)00669-9},
\newblock \eprint{hep-th/0005016}.

\bibitem{Chen:2020uac}
H.~Z. Chen, R.~C. Myers, D.~Neuenfeld, I.~A. Reyes and J.~Sandor,
\newblock \emph{{Quantum Extremal Islands Made Easy, Part I: Entanglement on
  the Brane}},
\newblock JHEP \textbf{10}, 166 (2020),
\newblock \doi{10.1007/JHEP10(2020)166},
\newblock \eprint{2006.04851}.

\bibitem{Hartman:2013mia}
T.~Hartman,
\newblock \emph{{Entanglement Entropy at Large Central Charge}}  (2013),
\newblock \eprint{1303.6955}.

\bibitem{Lin:2023ajt}
J.~Lin, Y.~Lu and Q.~Wen,
\newblock \emph{{Cutoff brane vs the Karch-Randall brane: the fluctuating
  case}},
\newblock JHEP \textbf{06}, 017 (2024),
\newblock \doi{10.1007/JHEP06(2024)017},
\newblock \eprint{2312.03531}.

\bibitem{Basu:2023wmv}
D.~Basu, J.~Lin, Y.~Lu and Q.~Wen,
\newblock \emph{{Ownerless island and partial entanglement entropy in island
  phases}},
\newblock SciPost Phys. \textbf{15}(6), 227 (2023),
\newblock \doi{10.21468/SciPostPhys.15.6.227},
\newblock \eprint{2305.04259}.

\bibitem{Chandra:2024bkn}
A.~Chandra, Z.~Li and Q.~Wen,
\newblock \emph{{Entanglement islands and cutoff branes from path-integral
  optimization}},
\newblock JHEP \textbf{07}, 069 (2024),
\newblock \doi{10.1007/JHEP07(2024)069},
\newblock \eprint{2402.15836}.

\bibitem{Akal:2020wfl}
I.~Akal, Y.~Kusuki, T.~Takayanagi and Z.~Wei,
\newblock \emph{{Codimension two holography for wedges}},
\newblock Phys. Rev. D \textbf{102}(12), 126007 (2020),
\newblock \doi{10.1103/PhysRevD.102.126007},
\newblock \eprint{2007.06800}.

\bibitem{Wen:2018mev}
Q.~Wen,
\newblock \emph{{Towards the generalized gravitational entropy for spacetimes
  with non-Lorentz invariant duals}},
\newblock JHEP \textbf{01}, 220 (2019),
\newblock \doi{10.1007/JHEP01(2019)220},
\newblock \eprint{1810.11756}.

\end{thebibliography}

	
	
	
		
		

	\end{document}